\documentclass[11pt]{JHEP3}
\usepackage{epsfig}
\preprint{ {\tt hep-th/0504042} \\ {EMPG--05--05} \\ {HWM--05--04}}
\newcommand{\be}[1]{ \begin{equation}\label{#1} }
\newcommand{\ee}{\end{equation}}
\newcommand{\bea}[1]{\begin{eqnarray}\label{#1} }
\newcommand{\eea}{\end{eqnarray}}
\newcommand{\eq}[1]{(\ref{#1})}

\newcommand{\fig}[1]{fig.\,\ref{#1}}
\title{Spacetime Properties of ZZ D-Branes}
\author{Michele Cirafici$^a$,
Justin R. David$^b$ , Edi Gava$^{b, c}$, K.S. Narain$^{b}$ \\
$^a$Department of Mathematics, \\  Heriot--Watt University, Riccarton, Edinburgh EH14 4AS, UK \\
$^b$High Energy Section, \\
 The Abdus Salam International Centre for Theoretical Physics,
\\Strada Costiera, 11-34014 Trieste, Italy.\\
$^c$Instituto Nazionale di Fisica Nucleare, sez. di Trieste, \\
and SISSA, Italy. \\
\email{michele@ma.hw.ac.uk, \\
justin, gava, narain@ictp.trieste.it} }

\abstract{
We study the tachyon and 
the RR field
sourced by the $(m,n)$ ZZ D-branes
in type 0 theories using 
three methods. 
We first use the 
mini-superspace approximation of the closed string wave functions 
of the tachyon and the RR scalar to probe these fields. 
These wave functions are then extended beyond the mini-superspace 
approximation using mild assumptions which are motivated by the 
properties of the corresponding wave functions in the mini-superspace
limit. These are then used to probe the tachyon and the RR
field sourced. 
Finally we study the space time fields sourced by the $(m,n)$ ZZ
D-branes using the FZZT brane as a probe. In all the three methods
we find that the tension of the $(m,n)$ ZZ brane is $mn$ times the 
tension of the $(1,1)$ ZZ brane. The RR charge 
of these branes is non-zero only for 
the case of both $m$ and $n$ odd,  in which case it is identical 
to the charge of the $(1,1)$ brane. As a consistency check we 
also verify that the space time fields sourced by the branes 
satisfy the corresponding equations of motion.
}

\begin{document}
\baselineskip 4ex
\section{Introduction}

The formulation of two dimensional string theories in terms of
matrix quantum mechanics has been considered as an example of
open/closed string duality. Recent developments in understanding
Liouville theory include 
the discovery of  D-brane 
states in Liouville theory which are localized at the strong coupling region
\cite{Zamolodchikov:2001ah},
and the subsequent study of the dynamics of these D-branes 
\cite{McGreevy:2003kb,Klebanov:2003km,McGreevy:2003ep,Martinec:2003ka,
Alexandrov:2003nn, Alexandrov:2003un}. This has given rise to evidence  
to place this duality along with the list of 
familiar open/closed string dualities 
obtained from the AdS/CFT correspondence.
Let us consider the 
bosonic two dimensional string theory. 
The $U(N)$ matrix quantum mechanics 
for this theory arises as a world volume theory of
the $N$ $(1,1)$ ZZ branes localized in the strong coupling region. 
The open string degrees of freedom on the $(1,1)$  branes are 
the open string tachyon and a gauge boson. The 
slope of the inverted 
harmonic oscillator potential of the matrix quantum 
is given by the mass of the tachyon, and the presence of the
gauge boson restricts the theory to the singlet sector.
Liouville theory also contains $(m,n)$ ZZ branes with $m\neq 1, n
\neq 1$ \cite{Zamolodchikov:2001ah}, 
the role of these branes in this duality is not
clear. These branes contain many more tachyons than the $(1,1)$ brane
and therefore one would expected them to be more unstable. 
An immediate question one can ask about the properties of these
branes is, what is their tension compared to that of the $(1,1)$ brane.
To our knowledge this question has not been addressed in the literature.

The $(m,n)$ ZZ branes also occur in 
two dimensional type 0 theories. These theories 
admit a holographic descriptions in terms of matrix quantum 
mechanics \cite{Douglas:2003up}. 
To be definite consider type 0B theory, the only two
physical fields for generic momenta  are the tachyon 
$T$  in the NS-NS sector and the RR scalar $C$. The matrix quantum
mechanics for this case arises as the world volume theory of unstable
$(1,1)$ branes.
Consider the D-instanton of this theory, this is a $(1,1)$ brane with 
Dirichlet boundary conditions  in the time direction, it carries 
a unit $RR$ scalar charge. One can similarly construct the $(m, n)$
D-instanton for $m \neq 1$ and $n\neq 1$. 
Since the long range fields of type 0B theory is just
the tachyon and the RR scalar, this D-instanton must source these
target space fields. 
Again, to elucidate the properties of the $(m,n)$ branes 
one can ask the question what is the behaviour of  
the long range closed string fields 
sourced by these branes and, can one  
distinguish the  index $(m,n)$ by 
studying these closed string fields.
In this paper we  address this question by 
evaluating the space time fields sourced by these branes.

The direct method to find the behaviour of a massless closed string 
field sourced
by a boundary state is to perform the following 
transform of the one point function of the
respective closed string operator on the boundary state.
\be{onepttr}
\Psi(z) = \int d^d p K(z, p) \frac{1}{p^2} \langle 0 |V_p |B\rangle 
\ee
Here $|B\rangle$ refers to the boundary state, and $\langle0| V_p$
refers to the closed string field. 
$K(z,p)$ is the wave function of closed string state which are eigen
states of  momentum $p$, $z$ stands for the target space coordinate. 
Evaluating the above transform
for the overlap of  Dp-brane boundary state  with say the 
graviton of critical string theories gives rise to the 
standard Coloumb behaviour $1/|z_\perp|^{d_\perp}$, where
$z_\perp$ refers to the transverse distance from the 
Dp-brane and $d_\perp$ is the number of spatial dimensions 
transverse to the brane. 
To perform the above transform for the case of the $(m, n)$
D-instanton we face two difficulties: the 
overlap $\langle 0|V_p |(m,n) \rangle$ has non-trivial 
momentum dependence, for the case of Dp-brane in critical string
theory the momentum dependence is just $\delta( p_\parallel)$
where $p_\parallel$ refers to the momentum along the 
Dp-brane.  The second difficulty is that unlike the 
case of critical string theory where $K(p, z) = e^{ip\cdot z}$ 
this wave function is at present not known. 

In this paper we use three methods to study the 
behaviour of  the closed string
field $\Psi(z)$  sourced by the $(m,n)$ D-instanton in type 0B theory.
We first use the mini-superspace approximation of
the wave-function $K(p,z)$ of the closed string tachyon $T$ and
the RR scalar $C$ of type 0B to study the 
behaviour fields sourced by the $(m,n)$ D-instanton. 
We find that the tension of an $(m,n)$ brane is proportional to 
$mn$ and the RR scalar charge of these branes is
non zero only for the case of $m$ and $n$ both odd and is equal
to that of the $(1,1)$ brane in the mini-superspace approximation. 
We then 
postulate the existence of these wave-functions  beyond 
the mini-superspace approximation. 
We make mild assumptions of the properties of the 
exact wave functions
which are motivated by the properties of the corresponding 
functions in the mini-superspace limit.
In this  approximation  
the wave-functions behave as a superposition of an incoming and an
outgoing wave with a relative reflection coefficient
far away from the Liouville potential. We assume that 
this feature still holds beyond the mini-superspace approximation 
with the reflection coefficient replaced by the exact 
reflection coefficient.  We also make the assumption that
the exact wave-functions can be written as a integral 
transform similar to that of the wave-functions in the 
mini-superspace limit.
With these assumptions we find that behaviour of the tachyon 
$T$ and the RR scalar $C$ sourced by the $(m,n)$ D-instanton 
in the region far away from from the Liouville potential
is identical to that of the behaviour obtained in the
mini-superspace approximation and they satisfy
the appropriate equations of motion.
Finally we study the behaviour of $\Psi(z)$ using the
FZZT brane \cite{Fateev:2000ik,Teschner:2000md}
of Liouville theory as a probe following
\cite{Kutasov:2004fg}. 
The prescription to find 
the closed string fields $\Psi(z)$ is as follows: the FZZT-ZZ annulus
amplitude $Z(m,n|\sigma)$ is written as Laplace transform of 
the field $\Psi(z)$
\be{sieb}
Z(m,n|\sigma) = \int_0^\infty \frac{dz}{z} e^{-z\cosh\pi \sigma} \Psi(z),
\ee
$z$ labels the target space coordinate.  Therefore one can 
find $\Psi(z)$ by performing the appropriate inverse Laplace
transform.  
Using this approach we  find that 
again the tension of an $(m,n)$ brane is proportional to 
$mn$ and the RR scalar charge of these branes is
non zero only for the case of $m$ and $n$ both odd and is equal
to that of the $(1,1)$ brane. 
Thus we see that all the three approaches 
yield the same results.
The fact that the RR charge of the $(m,n)$ D-instanton is not
proportional to the tension suggests that for $m\neq 1$ and $n\neq1$,
these branes are not stable
which can also be seen from the presence of many tachyonic modes in 
the open string spectrum.

This paper is organized as follows. In section 2. we review some 
properties of  ${\cal N}=1$ Liouville theory, 
the mini-superspace wave functions, 
the $(m, n)$ ZZ boundary states and the effective target 
space action of this theory. 
In section 3. we find the target space fields in the 
mini-superspace
limit and in section 4. we evaluate them again
by postulating the existence of the wave functions $K(p, z)$ beyond
the mini-superspace approximation satisfying mild assumptions. 
Finally in section 5. we use the FZZT probe method
study the target space fields. In all these approaches we find 
that the tension of the $(m,n)$ D-instanton is proportional
to $mn$ times the tension of a $(1,1)$ brane and the 
RR scalar charge of these branes is
non zero only for the case of $m$ and $n$ both odd and is equal
to that of the $(1,1)$ brane.

\section{Semi-classical wave-functions and ZZ boundary states} 

In this section we review some facts concerning the 
${\mathcal N}=1$ Liouville theory which will be used 
in the computations of the following sections.
In section 3 we will perform a semi-classical analysis of the
target space field configuration produced by a ZZ brane in type
0B string theory. For that purpose we need to recall  the
semi-classical wave-functions discussed in  
\cite{Douglas:2003up}. Recall that the Liouville theory
is characterized by a background charge $Q=b+1/b$, in terms of
which the central charge is $\hat{c}_L=1+2Q^2$. Together 
with a free
$\hat{c}_T=1$ system describing the time direction, this gives 
the correct value  $\hat{c}=10$ if $b=1$ \footnote{We work with
$\alpha'=1$.}. 
The semi-classical limit
corresponds to $b\rightarrow 0$, so we are actually dealing with a strongly
coupled world-sheet theory. Nevertheless, it will be instructive to
formally consider the semi-classical limit first. In that limit
the ${\cal N}=1$ sigma model reduces, in the Ramond sector,
to a supersymmetric quantum mechanical system. Due to supersymmetry,
the Hamiltonian has a doubly degenerate spectrum of wave-functions
for non-zero eigenvalue
$E=p^2$, $\psi_{p \pm}(z)$ where, if $\phi$ is the Liouville field and 
 $\mu$ the cosmological constant (which we take to
be positive from now on),   $z=\mu e^\phi$ and $p$ is 
the Liouville momentum.   
The wave-functions obey:
\begin{eqnarray} \label{1orderz}
    \left(z \frac {\partial} {\partial z} + z \right)
\Psi_{p+} (z) &=&  p \Psi_{p-} (z) \cr
    \left(- z \frac{\partial} {\partial z} + z \right)
\Psi_{p-} (z) &=&  p \Psi_{p+} (z),
\end{eqnarray}
which are basically supersymmetry transformations,
and the eigenvalue equations:
\begin{equation} \label{2orderz}
\left(- \left( z \frac{\partial}{\partial z} \right)^2 \mp z + z^2
- p^2 \right) \Psi_{p \pm} (z) = 0.
\end{equation}
This eigenvalue equation is nothing but that the statement
that the full Hamiltonian, i.e. the sum of Liouville part and
the free, time part, vanishes on physical states.

The term linear in $z$  in \eq{2orderz} 
is due to presence of the fermionic
zero-mode bilinear term in the Liouville Hamiltonian. 
Requiring delta-function normalizability, one gets as normalized
solutions: 
\begin{eqnarray} \label{mssfunctionsr}
 \Psi_{p+}(z) &=&  { 2 \over
 \Gamma \left(-{ip}+ {1 \over 2}\right)} \left({\frac{\mu}{2}}
 \right)^{-{ip}} \sqrt z \left(K_{{ip} -{1\over
 2}}(z) + K_{{ip} + {1\over 2}}(z)  \right)\\
 \Psi_{p-}(z) &=&  { 2i \over
 \Gamma \left(-{ip}+ {1 \over 2}\right)} \left({\frac{\mu}{2}}
 \right)^{-{ip}} \sqrt z \left(K_{{ip} -{1\over
 2}}(z) - K_{{ip} - {1\over 2}}(z)  \right).
\end{eqnarray}
For the non-supersymmetric Neveu-Schwarz sector, 
there no-fermionic zero modes and as a result the linear term 
$\pm z$ in the Hamiltonian is missing in (\ref{2orderz}). The resulting
normalized wave function is: 
\begin{equation}\label{mssfunctionns}
\Psi_{p 0} (z) = {2\over
 \Gamma\left(-{ip} \right) } \left(\frac{\mu}{2}
 \right)^{-{ip}} K_{ip}(z) 
\end{equation}
which satisfies the differential equation:
\begin{equation} \label{2orderzns}
\left(- \left( z \frac{\partial}{\partial z} \right)^2 + z^2
- p^2 \right) \Psi_{p 0} (z) = 0.
\end{equation}

These wave functions are normalized in such a way that
for $\phi\rightarrow -\infty$ (i.e. $z\rightarrow 0$) 
they are superpositions of an incoming ($e^{ip\phi}$)
and outgoing wave  ($e^{-ip\phi}$), in the form
$\Psi\rightarrow e^{ip\phi} + S(p) e^{-ip\phi}$, where
the phase $S$ is the reflection coefficient.
There are reflection coefficients for R and NS 
wave-functions $S^R(p)$ and $S^{NS}(p)$, obeying
\begin{equation}\label{reflection}\Psi_{-p\pm}=\pm S^R(p) \Psi_{p\pm} , 
~~~~~~\Psi_{-p0}= S^{NS}(p) \Psi_{p0}.
\end{equation}
Where reflection coefficients are given by;
\begin{equation}\label{refcoeff}
S^R(p)=\left({\frac{\mu}{2}}\right)^{2ip} \frac{\Gamma(ip+1/2)}{\Gamma(-ip+1/2)} ,~~~~~~
S^{NS}(p)=\left(\frac{\mu}{2}\right)^{2ip}\frac{\Gamma(ip)}{\Gamma(-ip)}
\end{equation}

The boundary states corresponding to (Dirichlet) ZZ
branes  \cite{Fukuda:2002bv,Ahn:2002ev}
are labelled by integers $(m,n)$ characterizing
the NS(R) degenerate representations of the 
${\mathcal N}=1$ Liouville theory for  $m-n=\rm {even(odd)}$.
We will be interested in the related wave-functions, which
are the (unnormalized) disc one-point functions of
the NSNS and RR ground states respectively. These are nothing
but the coefficients relating Ishibashi states, labelled by 
the continuum momentum $p$,  to Cardy states, labelled by $(m,n)$,
and are basically determined by the modular transformation
properties of the corresponding characters $\chi_{m.n}$ from the open
to the closed channel:
 \begin{eqnarray} 
 \label{ZZNS} \Psi_{\! NS}(p;m,n) &=&
    2\sinh(\pi mp/b)\sinh(\pi np b)
        \left[\frac{\Gamma(1\!+\!ip b)\Gamma(1\!+\!ip/b)}%
                {(2\pi)^{1/2}\; (-ip)} 
\left(\frac{\mu}{2}\right)^{-ip/b}\right]
                \\
\label{ZZR}  \Psi_{\! RR}(p;m,n) &=&
    -i^{m+n}
    \,2\sin[\pi m(\frac{1}{2} \!+\!ip/b)]
\sin[\pi n(\frac{1}{2}\!+\!ip b)]\times\\
       & & \left[\frac{\Gamma(\frac{1}{2}\!+\!ip b)
\Gamma(\frac{1}{2}\!+\!ip/b)}%
                {(2\pi)^{1/2} } 
\left(\frac{\mu}{2}\right)^{-ip/b}\right]. 
\nonumber
\end{eqnarray}
In \eq{ZZNS} and \eq{ZZR} 
$\mu$ actually stands for the renormalized cosmological
constant 
\be{rencos}
\mu = \mu_0 \gamma\left(\frac{1+b^2}{2} \right),  \;\;\;\;\;\;\;
{\rm with } \;\;\;\;
\gamma(x)=\frac{\Gamma(x)}{\Gamma(1-x)}
\ee

The expressions in \eq{ZZNS} and \eq{ZZR} 
are exact CFT results, valid for any $b$.
Their mini-superspace limit is obtained taking $b\rightarrow 0$
with $p/b$ finite and identifying $\mu_0$ with the semi-classical
cosmological constant. 

In the case in which $b^2$ is a rational number, in particular for $b=1$,
that will be the case we will study in this paper, the parametrization
of degenerate states with the pair $(m,n)$, with $m,n$ arbitrary 
positive integers is redundant\footnote{We thank N .Seiberg for
pointing this out to us and for bringing to our attention
reference \cite{Seiberg:2003nm}, where degenerate representations 
and corresponding boundary states for
$b^2$ rational are  discussed.}, since, for example for $b=1$,
all pairs $(m,n)$ with the same value of $m+n$ will correspond to the same
degenerate field. Also, the structure of null states
is richer compared to the generic irrational case.
However, one can show\cite{Seiberg:2003nm}, 
that for $b=1$, the consequence is simply that 
the degenerate representations, and therefore
the ZZ boundary states,  are labelled by $(t,1)$ with 
$t$ any positive integer. In any case, we will continue in the
following, to adopt the naive $(m,n)$ notation, with the understanding
that in the final results of sections 3,4,and 5 $(m,n)$
must be specialized  to $(t,1)$.

In the mini-superspace limit,
$\Psi_{\! NS}(p;m,n)$  obey 
the same reflection relation as $\Psi_{p0}$,
whereas, for $\Psi_{\! RR}(p;m,n)$, the  semi-classical reflection
relation is the same as that of $\Psi_{p+}$ times a phase $(-)^{m+n}$.
\begin{eqnarray}\label{reflectionbs}\Psi_{\! NS}(-p;m,n)&=& S^{NS}(-p) 
\Psi_{\! NS}(p;m,n)\\ 
\Psi_{\! RR}(-p;m,n)&=&(-)^{m+n}S^{R}(-p) 
\Psi_{\! RR}(p;m,n).\nonumber
\end{eqnarray}
Notice that these wave-functions do not contain
any free parameter corresponding to the position of the D-brane
in the $\phi$ space. In fact, the D-brane is stuck at $\phi\rightarrow
\infty$, which, due to the linear dependence of the dilaton
on $\phi$, is the region of strong coupling.

\subsection{Target space fields}

The type 0B theory has only two physical fields for
generic momenta, the tachyon $T$
in the NSNS sector, and the RR scalar $C$. The effective
action is expected to a classical solution involving 
a time independent (closed string) tachyon background
\begin{equation}\label{background}
T(\phi,t) = T(\phi) = \mu e^{\phi} \,.
\end{equation}
together with a linear dilaton and a flat two dimensional 
metric.
The tachyon couples to the RR scalar through the action
\cite{Douglas:2003up,Gross:2003zz}
\begin{equation}
\frac{1}{8\pi}\int e^{-2 T}\; dC\wedge *dC \label{kinetic}
\end{equation}
The natural field strength associated with $C$ is $F = e^{- T}dC$.
The linearized equations of motion and the Bianchi identities for
$C$ are given by
\begin{eqnarray}
d \left( e^T F \right) &=& 0 \cr d \left( e^{-T} * F \right) &=& 0
\label{eom}
\end{eqnarray}
that in component reads
\begin{eqnarray}
 \left(-{\partial \over \partial \phi} + \mu
 e^{\phi}\right)F_\phi &=& -{\partial \over \partial t} F_t \cr
 \left({\partial \over \partial \phi}+ \mu
 e^{\phi}\right)F_t &=& {\partial \over \partial t} F_\phi \ .
 \label{eomcomponent}
\end{eqnarray}
 
Equations (\ref{eom}), for the zero energy, time independent 
case, become $(\pm z\frac{\partial}{\partial z}+z)F=0$, whose 
solutions are $e^{\mp z}$, i.e.: 
\begin{eqnarray}
F &=& e^{-T} dt , \cr F &=& e^T d\phi \
\end{eqnarray}
that are called the electric and magnetic solutions, respectively.
Sometimes it can be useful to introduce the field $\chi = e^{-T}
C$ that has a canonically normalized kinetic term
\cite{Douglas:2003up}; in terms of $\chi$ the field strength is $F
= d \chi + \chi d T$.

A completely equivalent description can be given by introducing
the dual scalar $\tilde{C}$ related to $C$ by
\begin{equation}
e^{-T} dC = F = * \tilde{F} = e^T * d \tilde{C}.
\end{equation}
The RR scalar $C$ couples to D-instantons. D-instantons 
are D-branes localized both in the (Euclidean) time direction $t$, and
in the Liouville direction $\phi$. The corresponding boundary state
is the tensor product of the boundary state in the $t$ direction, 
times the boundary state in the $\phi$ direction. We have seen
that for the latter case there is a 
supersymmetric generalization of
the ZZ boundary state of the bosonic Liouville theory,
and the corresponding
wave function has been given in \eq{ZZNS} and \eq{ZZR}. As remarked in 
the previous section, while the position of
the D-instanton in the time direction is an arbitrary  modulus,
in the Liouville direction the position is frozen in the
strong coupling region  $\phi\rightarrow
\infty$. One of the questions that one can ask is what is
the meaning of the $(m,n)$ labels in the ZZ boundary states
from the point of view of the target space 0B theory. Various 
arguments have been given to suggest that only the (1,1) 
boundary state consistently describes the target
space D-instanton, to which the RR scalar $C$ couples minimally. 
It is still an open problem to give an interpretation 
to the cases with $(m,n)\neq (1,1)$. This is the question
we would like to address in this paper.

\section{Target space fields in the minisuperspace limit}

In this section we would like to discuss the the following 
problem: given a ZZ boundary state (times the free,
time component part), what is the field 
configuration, both for RR and NS-NS cases, it produces 
in $(t,\phi)$ space-time.
In principle, it is clear what we have to do: given the 
boundary state wave function $\Psi(p;m,n)e^{iEt_0}$ 
and the wave function 
corresponding to the closed string state, $\Psi_p(z)e^{iEt}$,
we have to evaluate:
\begin{equation} \label{fieldC}
\int \mathrm{d}p  \mathrm{d}E\!\ (\Psi_p(z))^*e^{-iEt} 
\frac{1}{p^2+E^2} \!\ \left(
\Psi(p;m,n)e^{iEt_0} \right).
\end{equation}
Here $t_0$ is the arbitrary position of the boundary state in
the time direction. To simplify the  calculation, we can assume 
a uniform distribution of D-branes in the time direction and
integrate over $t_0$, so that we have: 
\begin{equation}\label{fieldC1}
\int \mathrm{d}p  \!\ (\Psi_{p}(z))^*
\frac{1}{p^2} \!\  \left( \Psi(p;m,n) \right).
\end{equation}  
In (\ref{fieldC}), the factor $\frac {1}{p^2}$ is the propagator of the
canonically normalized field $\chi$. This is the appropriate
propagator, since the one-point function is given by a disk
computation with one bulk vertex operator insertion. 
In the vertex operator corresponds to a RR state then 
it has to be in the $\left( - \frac{3}{2},
-\frac {1}{2} \right)$ or 
in the $\left( - \frac{1}{2}, -\frac{3}{2} \right)$
picture. 
Thus for  RR bulk vertex operator in \eq{fieldC1} 
evaluates the RR potential. On the other hand if the bulk vertex
operator insertion is in the NSNS sector, it has to be in the 
$(-1, -1)$ picture. Since we are assuming an uniform distribution of
D-instantons in the time direction the configuration is 
T-dual to a $(m, n)$ D0-brane of type 0A theory.

In the familiar case of D-branes in critical
string theory, where the wave functions are just
plane waves and $(p, E)$ is replaced by the
momentum transverse to the brane, the above 
procedure is well known to reproduce
closed string field configurations which solve the linearized 
equation of motion coming from the target space effective action. 
In our case the time component of the wave function is
still a plane wave, but the Liouville part involves 
the appropriate wave functions. For the boundary state
wave functions, we have the exact expression
which are given in  \eq{ZZNS} and \eq{ZZR},
but for the wave functions of the closed string states we
have just the appropriate wave-functions in the 
mini-superspace limit given in \eq{mssfunctionsr} and
\eq{mssfunctionns}. 
A full string theory computation beyond the mini-superspace
limit, would need the knowledge of the exact Hamiltonian
and the corresponding eigen functions, which is missing. 

In any case, we will find instructive to analyze the expression
(\ref{fieldC1})
in the mini-superspace limit $b\to 0$. Therefore, we will take
the $b \to 0$ limit in the boundary
state wave functions \eq{ZZNS} and \eq{ZZR}. Note
that, if we had put in (\ref{fieldC}) the full boundary
state wave functions, the integral would have 
been badly divergent in the UV.

Let us begin by evaluating the RR field $C$
produced  by a $(1,1)$--brane. The wave functions
corresponding to this state are $\Psi_{p\pm}(z)$, but 
in view of their 
reflection properties, and of that of $\Psi_{RR}$,
only $\Psi_{p+}$ contributes in the integral of (\ref{fieldC1}),
which becomes, after taking $b\to 0$ and setting $m=n=1$ in
$\Psi_{RR}(p;m,n)$:
\begin{equation}\label{11brane}
\int \mathrm{d}p \!\  \sqrt z \left(K_{{ip} -{1\over
 2}}(z) + K_{{ip} + {1\over 2}}(z)  \right) \frac {1}{p^2} \cosh (\pi p).
\end{equation}
Notice that the factor $\Gamma(\frac{1}{2}+ip)$ as well as a $p$-dependent
phase have been canceled between the two wave functions.
The resulting integral is an even function of $p$ and is understood to be
suitably regulated both in the in the IR and in the UV.
We are actually interested in the field strength, which is given by
$F = d \chi + \chi dT$. For the background (\ref{background}) and
a time independent RR scalar, this is equivalent to the operator
$z\frac{\partial}{\partial z}+z $ acting on the
integral (\ref{11brane}). Since the only dependence on $z$ is
in the wave function $\Psi_{p+}$, we can use equations 
(\ref{1orderz}). The result of this
differential operator acting on (\ref{11brane}) is just 
to produce one more factor of $p$ in the numerator 
and to convert $\Psi_{p+}(z)$ into $\Psi_{p-}(z)$.
In order to evaluate the resulting $p$ integral it is 
convenient to use the integral representation for the modified
Bessel function \cite{Abram}:
\begin{equation} \label{Bessel}
K_\nu (z) = \int_0^\infty \mathrm{d}x \!\ x^{\nu - 1}
e^{-\frac{z}{2} \left(x + \frac{1}{x} \right)}.
\end{equation}
From the above integral representation it is clear that
$K_\nu(z)=K_{-\nu}(z)$. 
After applying $z\frac{\partial}{\partial z}+z $ to
(\ref{11brane}), we are therefore led to evaluate the following
type of integral: 
\begin{equation} \label{11braneF}
i\sqrt{z} \int_{- \infty}^{\infty} \mathrm{d}p \int_0^{\infty} \frac
{\mathrm{d}x}{\sqrt{x}} \int_0^\infty \mathrm{d}\tau  \!\ p \!\
(\frac{1}{x}-1) e^{i p \ln x} e^{\pi p a} e^{- \tau
p^2}e^{-\frac{z}{2} \left(x + \frac{1}{x} \right)}
\end{equation}
where  $a = \pm 1$, due to  $\cosh
\pi p=\frac{e^{\pi p}+ e^{-\pi p}}{2}$ , 
and we have used a Schwinger
parameterization for $\frac{1}{p^2}$. The integral can be 
regularized as usual by cutting off the domain of 
$\tau$-integration.
The $p$-integral in (\ref{11braneF}), being Gaussian, can
be readily performed. The resulting expression is proportional
to:
\begin{equation} \label{11braneG}
\sqrt{z} \int_0^{\infty} \frac
{\mathrm{d}x}{\sqrt{x}} \int_0^\infty \mathrm{d}y A(x) e^{-\frac{A^2(x)}{4}y^2}
(\frac{1}{x}-1) e^{-\frac{z}{2} \left(x + \frac{1}{x} \right)},
\end{equation}
where we have changed integration variable from
$\tau$ to $y=1/\sqrt{\tau}$ and defined $A(x)=\ln x-i\pi a$.
The $y$-integral is Gaussian for those values of $x$
for which ${\rm Re}A^2< 0$. The idea is then to first cut-off
the $y$-integral, deform the $x$-integration contour in a region
where we can remove the cut-off, so that  the $y$-integral
becomes Gaussian.

Let us describe the resulting
$x$-integration contour:  it will include the region
on the positive real axis where ${\ln x}^2 $
is sufficiently large, say $x> R$\footnote
{As we will see the final result will be independent
on this value, thus we do not need to specify it.}, together with  
$\left[ 0, \frac{1}{R} \right]$. 
In the region $x \in \left[ 0, \frac{1}{R} \right] \cup \left[ R,
+\infty \right]$ the $y$-integral is  Gaussian and the
integration straightforward: the integral 
$\int dy A(x) exp{- \frac{A^2(x)}{4}y^2}$  equals (up to a factor 
$\sqrt{\pi}$)
$+1$ for $x\in [R, +\infty]$ and $-1$ for $x\in [0, \frac{1}{R}]$.
Let us begin with the $a=+1$ term in
(\ref{11braneF}). 
To compute the
full integral, one needs to give a prescription to handle the
region $\left[ \frac{1}{R},R \right]$.
We analytically continue the $x$ integral 
in the complex plane, by choosing
a contour as shown in \fig{contouru} ; since $a=+1$
\FIGURE{
\label{contouru}
\centerline{\epsfxsize=10.truecm \epsfbox{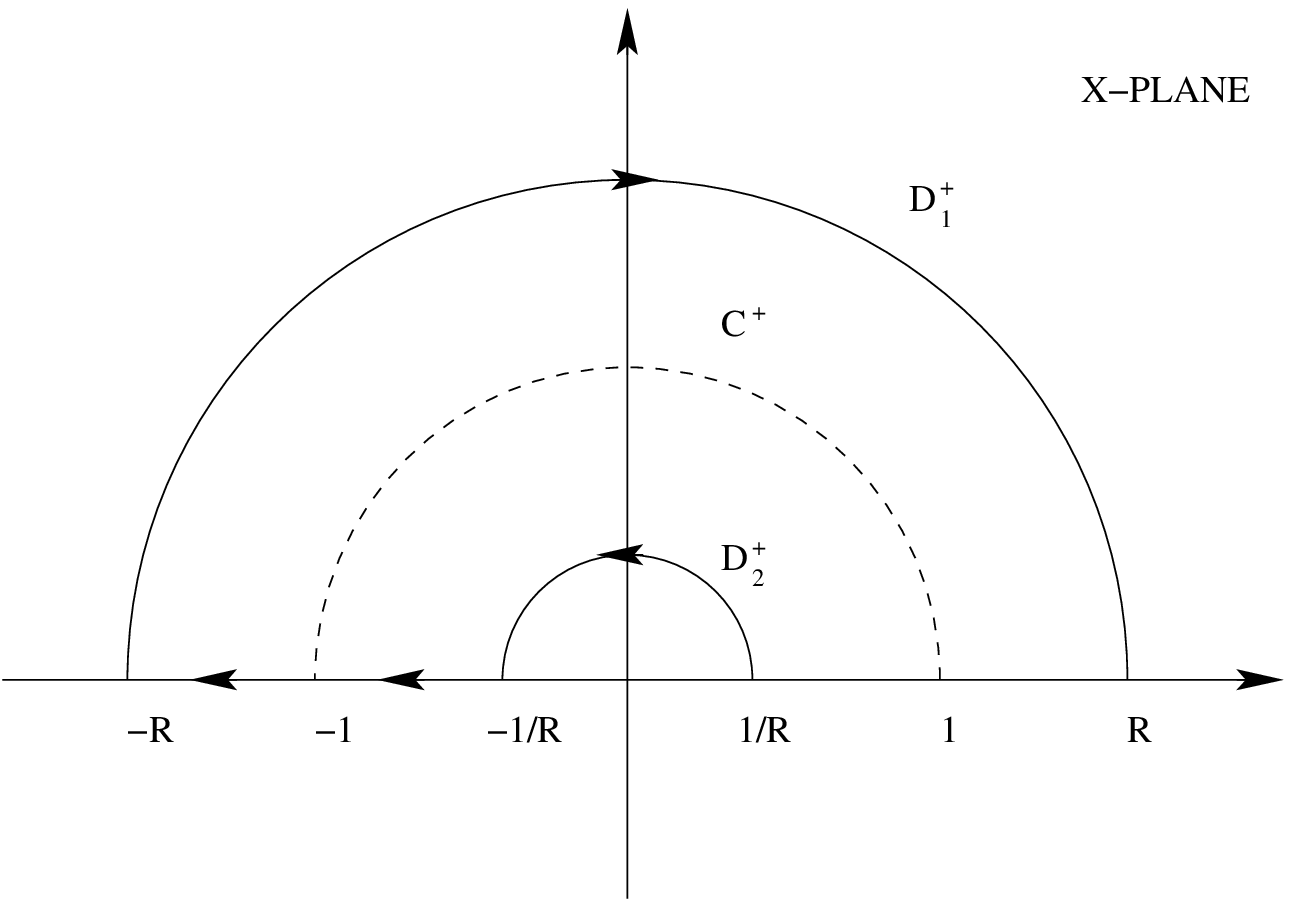}}
\caption{Contours in the upper half plane }
}
we close the contour in the upper half plane. The arcs
$\mathcal{D}_1^+$ and $\mathcal{D}_2^+$ are semi-circles of radius
$R$ and $\frac{1}{R}$ respectively, while $\mathcal{C}^+$ is the
unit semi-circle{\footnote {We write the superscript $+$ to remind
the reader that these contours are in the upper half plane}}. We
assume positive orientation clockwise. Thus for the $x$-integration
contour we have, schematically:
\begin{equation}\label{xcontour}
\int_{\frac{1}{R}}^{R} = - \int_{\mathcal{D}_2^+} +
\int_{-\frac{1}{R}}^{-1} + \int_{-1}^{-R} + \int_{\mathcal{D}_1^+}
\end{equation}
where the integrand is that of (\ref{11braneG}), with 
the $y$-integration still to be done.
At this stage we perform the $y$ Gaussian integral, the reason this 
can be done is that on the deformed contour the Gaussian integral 
is well defined. The integral: 
 $\int_0^\infty \mathrm{d}y A(x) e^{-\frac{A^2(x)}{4}y^2}$,
is proportional to $+1(-1)$ for $|x|>1 (|x|<1)$,
as a result, in (\ref{xcontour}) the first and
the second terms on the right-hand side flip sign. 
Now it is convenient to further split the integral
between $-\frac{1}{R}$ and $-R$  into 
two contributions, using the following relations:
\begin{eqnarray}\label{xcontour1}
\int_{-1}^{-R} + \int_{\mathcal{D}_1^+} &=& \int_{1}^{R} +
\int_{\mathcal{C^+}} \cr \int_{-1}^{-\frac{1}{R}} + \int_{\mathcal
{D}_2^+} &=& - \int_{\frac{1}{R}}^{1} + \int_{\mathcal{C^+}}
\end{eqnarray}
After substituting the above deformation of the contours we obtain
the contours shown in \fig{contouru1}.
\FIGURE{
\label{contouru1}
\centerline{\epsfxsize=10.truecm \epsfbox{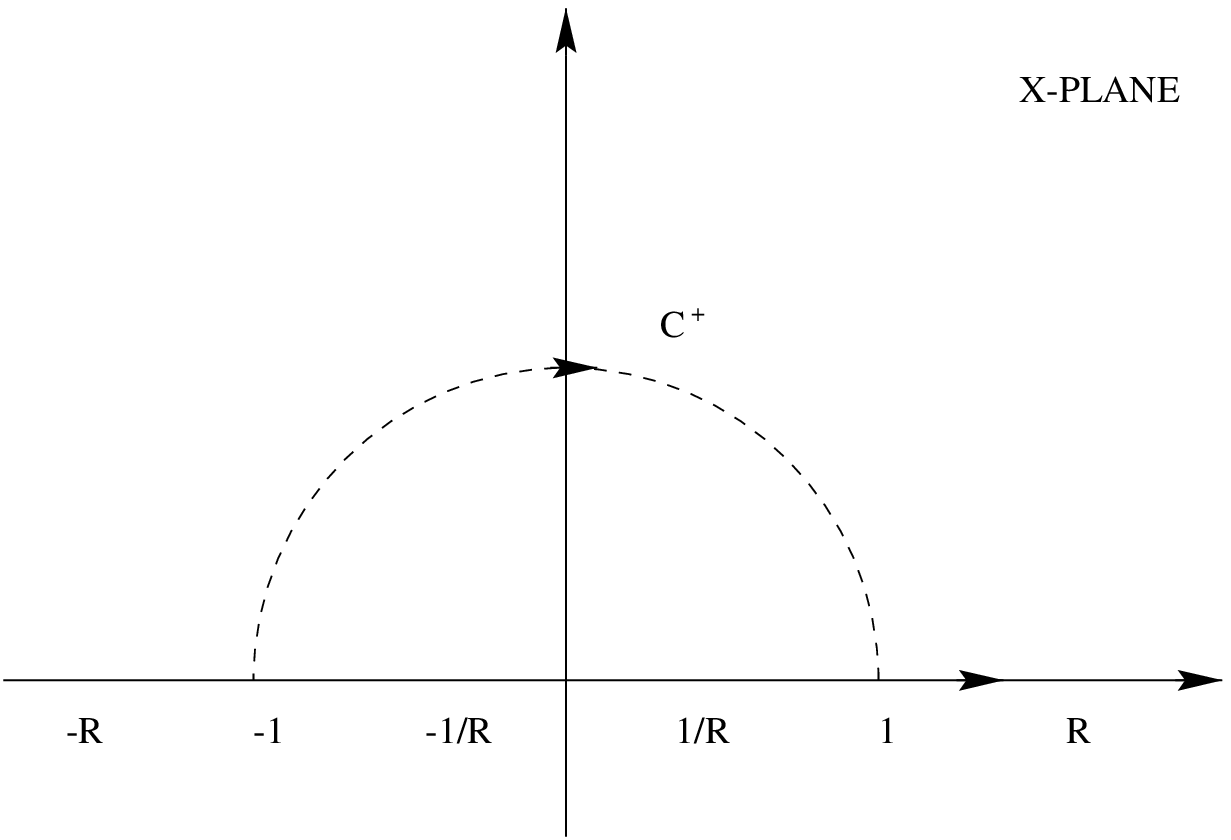}}
\caption{Final contour in the upper half plane }
}
Therefore,  using (\ref{xcontour1}) we can finally write  the equation 
\eq{11braneF}, for $a=+1$ as the $x$-integral: 
\begin{equation}\label{result11}
{\cal K} \left( 
\sqrt{z} \int_1^\infty \frac {\mathrm{d}x}{\sqrt{x}}
\left(\frac{1}{x} - 1 \right) \!\ e^{-\frac{z}{2} \left(x +
\frac{1}{x} \right)} - 2 \sqrt{z}  \int^\pi_0 \mathrm{d} \theta
\sin{\frac{\theta}{2}} e^{-z \cos{\theta}} \right)
\end{equation}
Here we have transformed the integral between $0$ and $1$ to an
integral between $1$ and $\infty$ by the change of variable $x \to
\frac{1}{x}$ and the second integral is the contribution from the
contour $\mathcal{C}^+$, parameterized as $e^{i \theta}$. ${\cal K}$ 
refers to the overall constant that we have not kept track of. Since
we are interested in comparing the target space fields of the 
$(m,n)$  brane to that of the $(1,1)$ brane it is not necessary to 
keep track of the overall constant.

To (\ref{result11}) we should now add the contribution coming from
the $a=-1$ term in (\ref{11braneF}). To compute this contribution
we follow the same pattern that led to (\ref{result11}), the only
difference being that, since $a=-1$ and $A(x)=
\ln x+i\pi$, the contour integrals close in
the lower half plane, as showed in  \fig{contourd}
\FIGURE{
\label{contourd}
\centerline{\epsfxsize=10.truecm \epsfbox{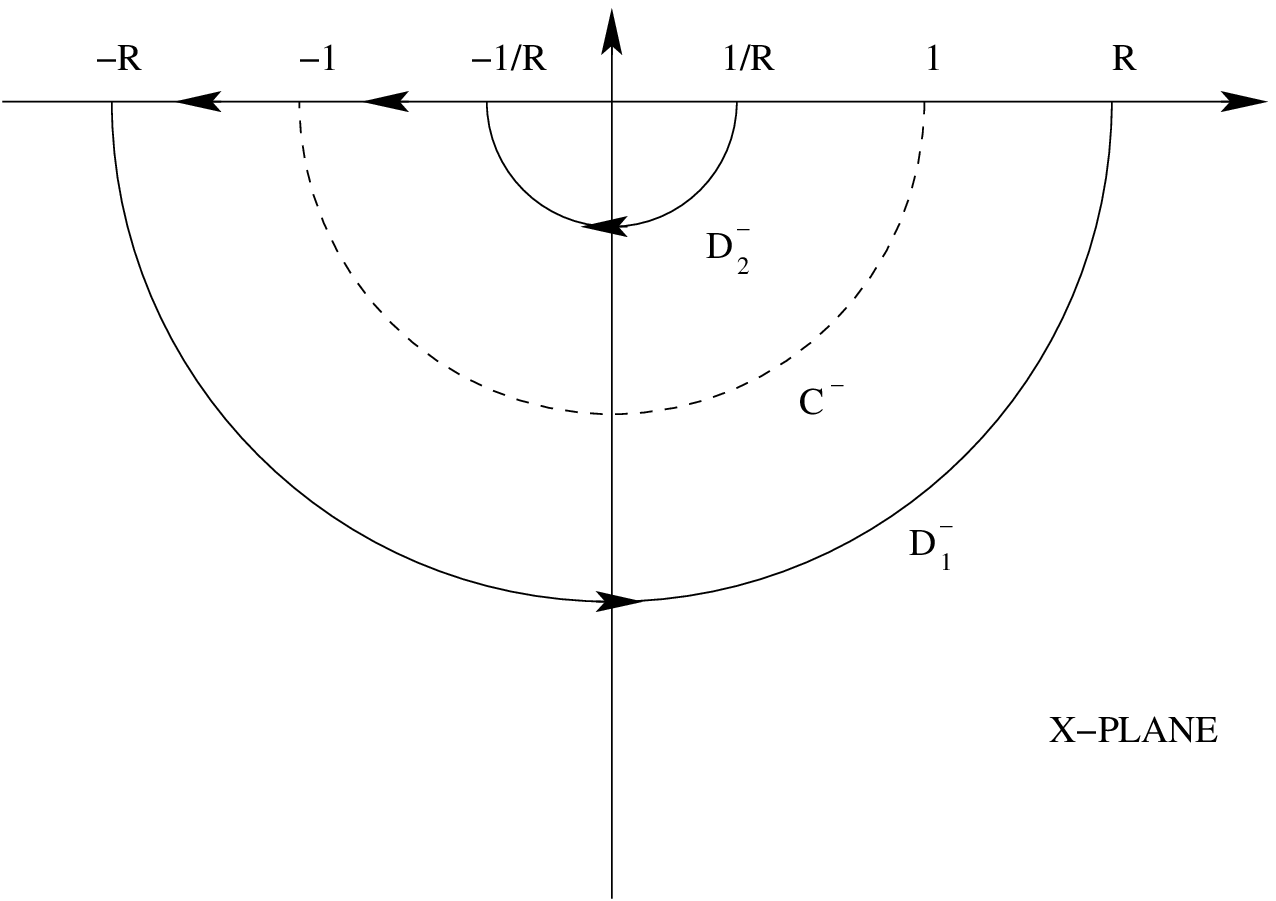}}
\caption{Contours on the lower half plane }
}
Now we introduce $\mathcal{D}_1^-$, $\mathcal{D}_2^-$
and $\mathcal{C}^-$ with radii respectively $R$, $\frac{1}{R}$ and
$1$, where the superscript $-$ points out that the contours close
in the lower half plane. These semi-circles are positively oriented
anti-clockwise. Thus we have
\begin{equation}
\int_{\frac{1}{R}}^{R} = - \int_{\mathcal{D}_2^-} +
\int_{-\frac{1}{R}}^{-1} + \int_{-1}^{-R} + \int_{\mathcal{D}_1^-}
\end{equation}
Again, now the integrals over $p$ and $\tau$ can be done, leading
us to
\begin{eqnarray}
\int_{-1}^{-R} + \int_{\mathcal{D}_1^-} &=& \int_{1}^{R} +
\int_{\mathcal{C^-}} \cr \int_{-1}^{-\frac{1}{R}} + \int_{\mathcal
{D}_2^-} &=& - \int_{\frac{1}{R}}^{1} + \int_{\mathcal{C^-}}
\end{eqnarray}
from which we obtain the same result as in (\ref{result11}). 
The final form of the contours are given in \fig{contourd1}.
\FIGURE{
\label{contourd1}
\centerline{\epsfxsize=10.truecm \epsfbox{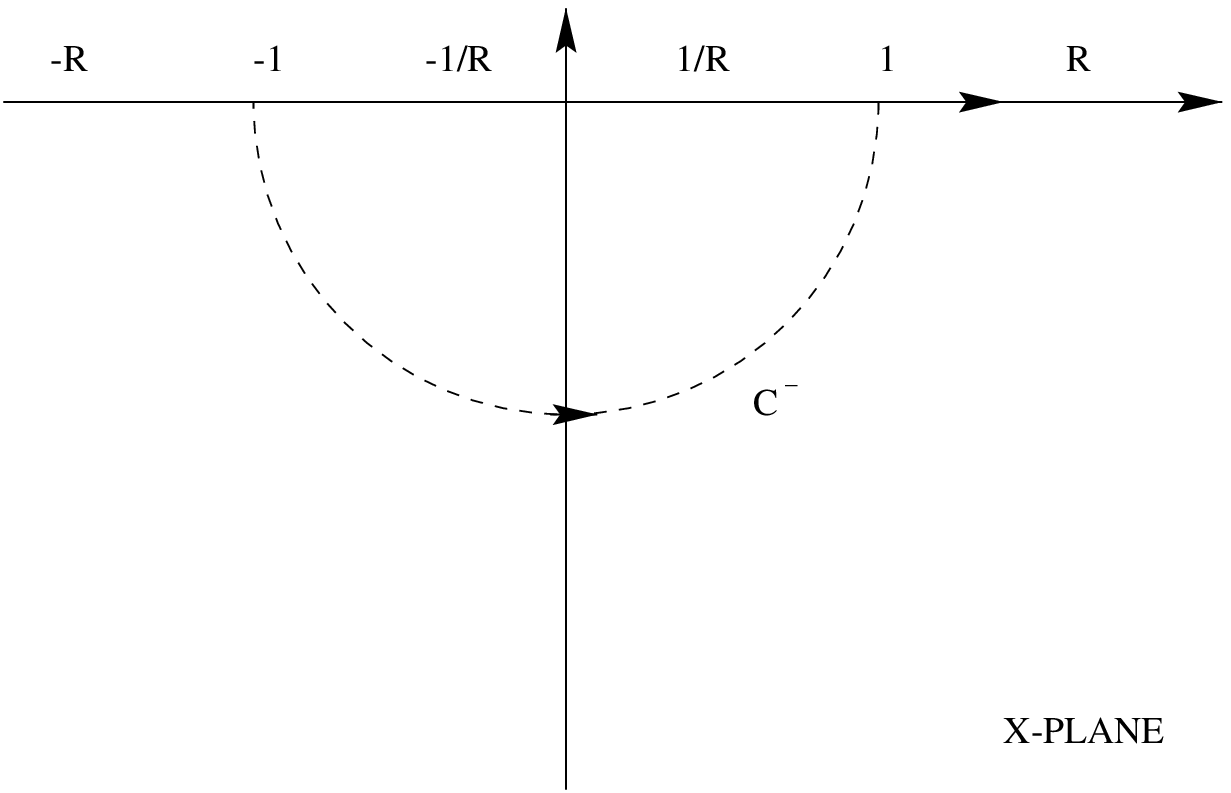}}
\caption{Final contour on the lower half plane }
}
Thus
the result of evaluating the integral (\ref{11braneF}) is
basically (\ref{result11}).

The formula \eq{result11} can in fact be simplified considerably.
Using the integral representation of the modified Bessel functions
$I_{1/2}$ and $I_{-1/2}$ \cite{Abram},  we have
\bea{inrepbes}
I_{1/2}(z)  + I_{-1/2}(z) &=& - \frac{1}{\pi} \left[
\int_0^\infty \frac{dx}{\sqrt{x}} \left( \frac{1}{x} -1 \right)
- 2\int_0^\pi   d\theta 
e^{z\cos\theta} \sin \frac{\theta}{2} \right], \\ \nonumber
&=& - \frac{1}{\sqrt{2 \pi z} } e^{z}
\eea
Substituting the above formula in \eq{result11} we see that it is
given by ${\cal K} e^z$. Thus the RR scalar field strength produced
by the $(1,1)$ ZZ brane is proportional to $e^{+z}$, which confirms
the analysis of \cite{Kutasov:2004fg} that the $(1,1)$ ZZ--branes
are localized at $\phi \to +\infty$. Also note that $e^z$ satisfies
the equation of motion as it satisfies the equation
\be{eomsi}
\left( z\frac{\partial}{\partial z} - z \right) e^{z} = 0
\ee
Therefore from  \eq{eomcomponent} we see that the $(1,1)$ D-instanton
distributed uniformly in the Euclidean time direction sources the
magnetic RR scalar.

In fact one can directly 
check that equation (\ref{result11}) satisfies the
equations of motion, this is useful to generalize to the case of 
$(m,n)$ brane. 
To do this one should apply the second
order differential operator (\ref{2orderz}) directly to the
potential (\ref{11brane}). This is equivalent to apply the first
order differential operator $z \frac{\partial}{\partial z}-z $
directly to the field strength given by (\ref{result11}). To check
this, the following formula has to be used:
\begin{eqnarray} \label{checkeom}
&& \left( z \frac{\partial}{\partial z} - z \right) \sqrt{z}
\int_1^\infty  \!\  \frac {\mathrm{d}x}{\sqrt{x}} \left(
\frac{1}{x} -1 \right)e^{-\frac{z}{2} \left(x + \frac{1}{x}
\right)} \cr &=& -  \sqrt{z} \int_1^\infty  \!\ \mathrm{d}x \frac
{\partial} {\partial x} \left(  \sqrt{x} \left( \frac{1}{x} + 1
\right) e^{-\frac{z}{2} \left(x + \frac{1}{x} \right)} \right) = 2
\sqrt{z} e^{-z}
\end{eqnarray}
Note that, due to the formula
above, for the equation of motion to be satisfied, only the
integration limits matter.
What happens is that after applying  $z \frac{\partial}{\partial z}-z $
the integrand becomes a total derivative in $x$ and the 
boundary contributions from the two terms in (\ref{result11}),
from $x=1$ and $\theta=0$, cancel each other. The fact
that  $z \frac{\partial}{\partial z}-z $ 
annihilates the field strength,  implies that 
the latter is proportional to $e^z$.
This coincides with the non-normalizable solution of the 
equations of motion discussed in \cite{Douglas:2003up}
and reviewed in section 2.

\subsection{$(m,n)$ branes : The RR  field}

In this subsection we will extend the previous discussion
to the case $(m,n)\neq(1,1)$. 
Let us take the mini-superspace limit  
of the RR wave functions (\ref{ZZR}). We have to distinguish
the cases of $n$ even and $n$ odd: in the case of $n$
even, one sees from (\ref{ZZR}) that  $\Psi_{RR}(p;m,n)$
gives a factor $npb$ for $b$ small and this results in
an extra power of $p$ in the numerator, compared to the
previous case, in the integrand for the field strength.
By regularizing the integral as before, one can verify that
in this case the second integration (in $\tau$) gives a
vanishing result. 
In the case of $n$ odd
$\Psi_{RR}(p;m,n)$ becomes, up to a phase,
the same as $\Psi_{RR}(p;m,1)$ in the $b\to 0$ limit.
The resulting integral will be evaluated along the same
lines indicated in the previous subsection for the (1,1) case.
We have in turn to consider separately the cases
of $m$ even and $m$ odd. In the first case, due to the reflection
properties of $\Psi_{p\pm}$ and $\Psi_{RR}(p;m,n)$ discussed
in section 2, we see that only $\Psi_{p-}$ can enter in the integral
of the type \ref{fieldC1}. Therefore we will need to evaluate:

\begin{equation} \label{1mbranemeven} i^{m+n+1}\int \mathrm{d}p \!\
\sqrt{z}(K_{\frac{1}{2}+ip}(z)-K_{\frac{1}{2}-ip}(z))
\frac {1}{p^2} \sinh[\pi m p], \!\,~~~ m ~\rm{even}.
\end{equation}

For the same reasons,
for $m$ odd only $\Psi_{p +}(z)$ can contribute and we have
the integral:

\begin{equation}\label{1mbranemodd}
i^{m+n} \int \mathrm{d}p \!\
\sqrt{z}(K_{\frac{1}{2}+ip}(z)+ K_{\frac{1}{2}-ip}(z))  \frac {1}{p^2} 
\cosh[\pi m p],\!\,
~~~ m~{\rm odd}.
\end{equation}
Notice that again the $\Gamma$ functions and the $p$-dependent
phases cancel between  $\Psi_{RR}(p;m,n)$ and the normalization
of $(\Psi_{p\pm}(z))^*$.

Let us begin with the latter case, $m$ odd: 
we apply to (\ref{1mbranemodd}) the
operator (\ref{1orderz}) to obtain the field strength. We need 
then to evaluate integrals of the type:
\begin{equation} \label{1moddbraneF}
\int_{- \infty}^{\infty} \mathrm{d}p \int_0^{\infty} \frac
{\mathrm{d}x}{\sqrt{x}} \int_0^\infty \mathrm{d}\tau  \!\ p \!\
(\frac{1}{x}-1) e^{i p \ln x} e^{\pi m p a} e^{- \tau
p^2}e^{-\frac{z}{2} \left(x + \frac{1}{x} \right)}
\end{equation}
where the factor containing $a = \pm 1$ arises when writing $\cosh
\pi m p$ in terms of exponentials. We are facing the same convergence
problems we had in the case $m=1$ and we follow the same procedure.
Let us start with
$a=+1$. To compute this integral, we choose again to promote $x$
to a complex variable and deform the contour of integration
after having regulated the integral; but
this time, due to the integer $m>1$, the divergence 
in the $y$-integral is worse. To cure this problem, it is sufficient to
continue the semi-circles $\mathcal{D}$ and $\mathcal{C}$ to a
contour ending into the $\frac{m+1}{2}$-th sheet in the (infinite)
covering of the $x$ plane . In other words, we take a
``spiral'' that winds around the origin counterclockwise(clockwise)
with an angle  of $m\pi$, 
starting from a given point on the real $x$-axis when $a$ 
is positive(negative).
Then, the previous interval $(Re^{i\pi}, \frac{1}{R}e^{i\pi}))$
on the negative real $x$ axis,
is replaced now by  $(Re^{im\pi}, \frac{1}{R}e^{im\pi}))$, with
some different $R$.
This is exactly the angle needed to undo
the  term $-im\pi$ in $ \ln x
- i m \pi$ in order to compute the Gaussian $y$-integral. We
then proceed with the $x$-contour deformation argument following
the same steps outlined for the (1,1)--brane, arriving at:
\begin{equation}\label{resultm1modd}
\sqrt{z} \int_1^\infty \frac {\mathrm{d}x}{\sqrt{x}} \left(
\frac{1}{x} - 1 \right) \!\ e^{-\frac{z}{2} \left(x + \frac{1}{x}
\right)} - 2 \sqrt{z} \int^{m \pi}_0 \mathrm{d} \theta
\sin{\frac{\theta}{2}} e^{-z \cos{\theta}}
\end{equation}
The same procedure is  utilized when $a=-1$, leading again to
(\ref{resultm1modd}). Moreover, one can check, by using formula
(\ref{checkeom}), that the equations of motions are satisfied.
The reason is that, due to (\ref{checkeom}), only the
boundary terms matter and in (\ref{resultm1modd}) the boundary term
is $1$ in the first integral and in $0$ in the second cancel each
other, 
while the others vanish.

Let us look more closely at (\ref{resultm1modd}): 
The $m$ dependence is
rather peculiar; in fact the second integral obeys the following
property
\begin{eqnarray}
\int_0^{\pi} \mathrm{d} \theta \sin{\frac{\theta}{2}} e^{-z
\cos{\theta}} = \int_{\pi}^{2 \pi} \mathrm{d} \theta
\sin{\frac{\theta}{2}} e^{-z \cos{\theta}} = \cr = - \int_{2
\pi}^{3 \pi} \mathrm{d} \theta \sin{\frac{\theta}{2}} e^{-z
\cos{\theta}} = - \int_{3 \pi}^{4 \pi} \mathrm{d} \theta
\sin{\frac{\theta}{2}} e^{-z \cos{\theta}}
\end{eqnarray}
that can be proven by using the properties of the trigonometric
functions and the fact that the integrand is a periodic function
with period $4 \pi$. This means that, in (\ref{resultm1modd}),
where $m$ is odd, if we split the second integral in $m$ integrals
over intervals of length $\pi$, all the terms pairwise cancel
except one, leading us back to the result (\ref{result11}). It
seems that, at least in the mini-superspace approximation,
$(m,n)$-branes with $m$ and $n$ odd, behave exactly as (1,1) branes with
respect to the RR field, {\it i.e.} they have the same RR charge. 
In fact, just as for the case of the $(1,1)$ brane the result of the
integral in \eq{resultm1modd} is just proportional to $e^z$. Therefore
it is clear that the RR field sourced by the $(m,n)$ brane satisfies 
the equation of motion \eq{eomsi}.

Let us suppose now that $m$ is an even integer. The two
contributions arising from (\ref{1mbranemeven}), when computing the
field strength are of the form:
\begin{equation} \label{1moddbraneF1}
i^{m+n}\int_{- \infty}^{\infty} \mathrm{d}p \int_0^{\infty} \frac
{\mathrm{d}x}{\sqrt{x}} \int_0^\infty \mathrm{d}\tau  \!\ p \!\
(\frac{1}{x}+1) e^{i p \ln x} e^{\pi m p a} e^{- \tau
p^2}e^{-\frac{z}{2} \left(x + \frac{1}{x} \right)}
\end{equation}
The differences with respect to the $m$--odd case, are the sign in
the factor $1+\frac{1}{x}$ and the fact that now the contributions
with $a=\pm$ carry a different relative sign that comes directly
from $\sin m \pi p$. The integral can be computed using the same
procedure which has been detailed for the case 
of $m$ odd.
Again, after
doing the $p$ and $\tau$ integrals, one finds three contributions:
\begin{equation}
\sqrt{z} \left( - \int_0^1  + 2 \int_{\mathcal{C}} +
\int_{1}^{\infty} \right)  \frac {\mathrm{d}x}{\sqrt{x}} \left(
\frac{1}{x} + 1 \right) \!\ e^{-\frac{z}{2} \left(x + \frac{1}{x}
\right)}
\end{equation}
This time, due to the factor $\frac{1}{x}+1$ the first and the
last integral cancel each other. What remains is:
\begin{equation}
2 i \sqrt{z} \int_{0}^{m \pi} \mathrm{d} \theta
\cos{\frac{\theta}{2}} e^{-z \cos{\theta}}
\end{equation}
But now one can show that this integral vanishes. In fact,
due to the periodicity of the integrand, the relevant integral to
compute is just the one from $0$ to $2 \pi$. But this integral
vanishes since
\begin{eqnarray}
&& \int_{0}^{\pi} \mathrm{d} \theta \cos{\frac{\theta}{2}}
e^{-z \cos{\theta}} = \int_{\pi}^{2 \pi} \mathrm{d} \theta
\sin{\frac{\theta}{2}} e^{z \cos{\theta}} = \cr &&= -
\int_{\pi}^{2 \pi} \mathrm{d} \theta'
\cos{\frac{\theta'}{2}} e^{-z \cos{\theta'}}
\end{eqnarray}
where $\theta' = -\theta + 3 \pi$.
The same computation carries on for the case of $a=-1$ with
exactly the same vanishing result.
Combining all the cases, 
we find that  that $(m,n)$-branes 
are decoupled from the RR case except when   
both $m$ and $n$ are odd. For this case, the RR charge is identical to
that of the $(1,1)$ brane.

\subsection{$(m,n)$ branes: The NS-NS field}

We proceed now to discuss the NSNS field produced
by $(m,n)$ ZZ branes, which, in particular, contains the information
about their tension. 
We have the wave-function corresponding to the NSNS
ground state (\ref{mssfunctionns}) in the semi-classical limit and
the exact disc NSNS one-point function, (\ref{ZZNS}). In the $b\to 0$ limit
the leading term in the latter is of order $b$ and proportional
to $n$. More precisely, after taking into account of a $1/b$
factor from the $p$-measure, equation (\ref{fieldC1}) is proportional
to
\begin{equation}\label{1mbrane}
n \int \mathrm{d}p \!\ K_{ip}(z) \frac {1}{p^2} \sinh(\pi
mp)p, 
\end{equation}
where the factor $np$ comes from the $b \to 0$ 
limit of (\ref{ZZNS}). 
Here the  $p$
appearing in (\ref{1mbrane}) is the $p$ of (\ref{ZZNS})
divided by $b$.  Note that the only $n$ dependence is a
multiplicative factor outside the integral. The integral
(\ref{1mbrane}) can be reduced to integrals of the kind
\begin{equation}
n \int \mathrm{d}p \int_{0}^{\infty} \mathrm{d}t \int_{0}^{\infty}
\frac{\mathrm{d}x}{x} e^{i p \log x} e^{\pi m p a} p e^{-t p^2}
e^{-\frac{z}{2} \left( x + \frac{1}{x} \right)}
\end{equation}
where $a$ can be $\pm 1$ and the exponential containing $a$ arises
from the $\sinh(\pi mp)$. We can evaluate this integral by following the
method discussed earlier. We arrive at:
\begin{equation}
- ni \int_{0}^{1} \frac{\mathrm{d}x}{x} \!\ e^{-\frac{z}{2} \left(
x + \frac{1}{x} \right)} + ni \int_{1}^{\infty}
\frac{\mathrm{d}x}{x} \!\ e^{-\frac{z}{2} \left( x + \frac{1}{x}
\right)} + 2 ni \int_{\mathcal{C}} \frac{\mathrm{d}x}{x} \!\
e^{-\frac{z}{2} \left( x + \frac{1}{x} \right)}
\end{equation}
The first two terms cancel each other due to the symmetry
properties of the integrand under the transformation $x \to
\frac{1}{x}$; the last term is integrated along the contour
$\mathcal{C}$ that winds around the origin covering an angle of $m
\pi$ and can thus be written as:
\begin{equation}\label{nsfield}
 2 m n  \int_{0}^{\pi m} \mathrm{d} \vartheta e^{- z \cos
 \vartheta}  = {\cal K}  m n I_0 (z)
\end{equation}
where $I_0(z)$ is the (modified) Bessel function \cite{Abram} and ${\cal K}$
refers to the normalization constant which we have not kept track of.
Note that $I_0(z)$ is annihilated by the operator
$(-(z\frac{\partial}{\partial z})^2+z^2)$, 
which is the mini-superspace Hamiltonian (for zero energy)
in the NS sector, i.e. equation (\ref{2orderzns}) for $p=0$
and without the linear term in $z$. In fact $I_0$
is the exponentially growing solution of this
differential equation, $I_0 \sim e^z/\sqrt{z}$ for large $z$. 
This result is in agreement with what was found for the
RR case, where also the field was exponentially growing
at large $z$.
Finally, (\ref{nsfield}) indicates that the tension 
of $(m,n)$ branes, at least in
the mini-superspace approximation is proportional to the  product $mn$.

\section{Beyond the semi-classical approximation}

We have already stressed that for the case of
non-critical type 0B theory we are interested in,
the Liouville dynamics is far from the semi-classical
regime of small $b$ (or large $b$ as $b \rightarrow 1/b$ is a
symmetry)
since $b$ is actually at the self-dual point $b=1$.
What is missing, in order to extend the strategy
of the previous section beyond the semi-classical limit,
are the expressions for the ground state wave-functions.
In general we expect the equation of motions to
receive higher order corrections in $b$.   
Unfortunately, the
generalizations of 
the mini-superspace wave functions
\eq{mssfunctionsr} and \eq{mssfunctionns} are 
presently not known. 
We have however some  non-perturbative information about
the exact wave-functions. This is given by the
exact reflection coefficients both for RR and NSNS ground
states. In fact, they can be read-off 
from the expressions for the boundary states wave-functions,
\eq{ZZNS} and \eq{ZZR}.
Setting $b=1$, for the NSNS sector these are given by: 
\begin{equation} \label{reflNS}
{S}^{NS}(p) = -\left(\frac{\mu}{2}\right)^{-2ip}
\frac{{\Gamma}^2(ip)}{{\Gamma^2(-ip)}},
\end{equation}
and for the RR sector they are given by:
\begin{equation} \label{reflRR}
{S}^{RR}(p) =
\left(\frac{\mu}{2}\right)^{-2ip}
\frac{{\Gamma^2(\frac{1}{2}+ip)}}{{\Gamma^2(\frac{1}{2}-ip)}}.
\end{equation}
Here again we have denoted with $\mu$ what actually is the renormalized 
cosmological constant 
defined in \eq{rencos}.
The exact, quantum wave functions
obey reflection relations of the type (\ref{reflection}) with
$S^{NS,RR}$ given in \eq{reflNS} and \eq{reflRR}. 
The semi-classical wave-functions (\ref{mssfunctionns})
behave, in $z \to 0$ limit, far away from the Liouville
potential, as a superposition of an incoming and an outgoing wave with
a relative reflection coefficient. We assume that this feature 
still holds beyond the mini-superspace approximation, with the
mini-superspace reflection coefficients replaced by the
quantum version, \eq{reflNS} and \eq{reflRR}.
We then postulate the existence of ``quantum'' wave functions,
expressible in the form:
\begin{eqnarray} \label{psiNS}
\Psi^{NS}(p,z) &=& -\frac{(\mu/2)^{-ip}}{\Gamma^2(-ip)}
\int_{0}^{\infty} \mathrm{d}x \!\ x^{i p -1} 
f^{NS}(x,z) \\\label{psiRR} \Psi^{RR}_{\pm}(p,z) &=& 
 \frac{(\mu/2)^{-ip}}{\Gamma^2(\frac{1}{2}- i
p)}\int_{0}^{\infty} \mathrm{d}x \!\ x^{i p
-1} f^{RR}_{\pm}(x,z),
\end{eqnarray}
where the functions $f^{NS,RR}(x,z)$ obey the relations:
\begin{eqnarray} \label{xto1overx}
f^{NS}(x,z) &=& f^{NS}(\frac{1}{x},z) \cr f^{RR}_{\pm}(x,z) &=&
\pm f^{RR}_{\pm}(\frac{1}{x}, z)
\end{eqnarray}
that is similar to the properties of 
\eq{mssfunctionsr} and \eq{mssfunctionns}, that can be obtained from the
integral representation of the modified Bessel function
(\ref{Bessel}). 

The wave functions (\ref{psiNS})
and (\ref{psiRR}) should reduce in the weakly coupled region ({\it
i.e.} in the $z \to 0$) limit to a superposition of incoming and
outgoing waves, the relative phase being given by the
reflection coefficients (\ref{reflNS}) and (\ref{reflRR}),
respectively. These can be interpreted as conditions on the 
wave-functions $f^{NS,RR}(x,z)$. From these conditions one finds
the $z \to 0$ behavior of the functions $f^{NS,RR}(x,z)$
\begin{eqnarray}\label{effes}
f^{NS}(x,z) &\to_{z \to 0}&  \left(
K_0\left(
\sqrt{2xz} \right) - K_0\left(\sqrt{\frac{2z}{x}}\right) \right) \\
f^{RR}_{+}(x,z) &\to_{z \to 0}& \sqrt{z} 
\left( \sqrt{x} K_0\left(\sqrt{2x z}\right) +
\frac{1}{\sqrt{x}} K_0\left(\sqrt{\frac{2z}{x}}\right) \right)\\
f^{RR}_{-}(x,z) &\to_{z \to 0}&\sqrt{z} 
\left( - \sqrt{x} K_0\left(\sqrt{2x z}\right) +
\frac{1}{\sqrt{x}} K_0\left(\sqrt{\frac{2z}{x}}\right) \right)\\
\nonumber
\end{eqnarray}
The fact that these expression reproduce the right asymptotic behaviour
can be explicitly verified. This is done by performing the $x$-integral in
\eq{psiNS} and \eq{psiRR} with the use of the integral representation
for the modified Bessel function given in  
(\ref{Bessel})\footnote{One can easily generalize 
\eq{psiNS},  \eq{psiRR} and  \eq{effes} to the
case of $b\neq 1$. The subsequent analysis however becomes
more involved.}.

We can push further the analogy with the mini-superspace wave
functions (\ref{mssfunctionsr}), by demanding the analog of
equations (\ref{1orderz}) and (\ref{2orderz}). Namely, we postulate
the existence of differential operators ${D}^2_{NS}
(z,\frac{\partial}{\partial z})$ and 
${D}^{\pm}_{RR}(z,\frac{\partial}{\partial z})$
acting on the wave-functions  such that:
\begin{eqnarray} \label{1orderfull}
\mathcal{D}^+_{RR} \Psi_+^{RR} (z,p) &=& p \Psi_{-}^{RR}(z,p) \\
\nonumber
\mathcal{D}^-_{RR} \Psi_-^{RR} (z,p) &=& p \Psi_{+}^{RR}(z,p)
\end{eqnarray}
and
\begin{equation}\label{firstsecond}
\mathcal{D}^2_{RR} \Psi_{\pm}^{RR} (z,p) \equiv
\mathcal{D}^+_{RR}\mathcal{D}^-_{RR} \Psi_{\pm}^{RR} (z,p) = 0,
\end{equation}
as well as:
\begin{equation}\label{2orderfull}
\mathcal{D}^2_{NS} \Psi^{NS} (z,x) = 0
\end{equation}
More generally, we may assume that there is a $b$-dependent 
family of each of these operators, interpolating between
the mini-superspace expressions \eq{1orderz} and \eq{2orderz}
in the $b\to 0$ limit and some (unknown) expression for the case $b=1$.
On the other hand, the factorization
relation, (\ref{firstsecond}), is a consequence
of world-sheet supersymmetry, which is an exact symmetry,
independent of $b$. Notice that it is not
obvious that the differential operators  ${D}^2_{NS}
(z,\frac{\partial}{\partial z})$ and 
${D}^{\pm}_{RR}(z,\frac{\partial}{\partial z})$, 
remain of order two and one respectively. 
Nevertheless, we may expect that the  
the structure of equations \eq{1orderfull} and \eq{2orderfull}
is still valid, since this arises from supersymmetry and
from the fact the $t$-dependent part of the full
wave-function is in any case a plane-wave, $e^{ipt}$.
Regarding  \eq{psiNS} and \eq{psiRR},
the assumption that the wave-functions can be written
in that way seems rather natural, being just a
definition of their Fourier transforms. The property  
(\ref{xto1overx}), on the other hand, 
is a symmetry property obeyed by the corresponding
mini-superspace wave functions which we have assumed to 
be true beyond the $b\rightarrow 0$ limit.
From (\ref{psiRR}) and (\ref{1orderfull}) it follows, by a partial
integration, that   $f^{RR}_{\pm}(x,z)$ satisfies the following
identity:
\begin{equation}\label{eqmot}
\mathcal{D}^{\pm}_{RR}(z) f_{\pm}^{RR} (x,z) =i x\frac{\partial}
{\partial x} f_{\mp}^{RR}(x,z). 
\end{equation}
This equation generalizes the relation (\ref{checkeom})
beyond the mini-superspace limit, and  will be used when we 
will check the equation of motion for the RR field strength.

\subsection{The RR field}

Let us begin with the evaluation of the RR field strength. Again,
the idea is to write down an integral of the kind of
(\ref{11brane}) with the full one-point function and the wave
functions (\ref{psiNS}) and (\ref{psiRR}) (with the appropriate
parity, depending on $m$ and $n$ being even or odd). Generically we have
to evaluate:
\begin{equation}
\int \mathrm{d}p \!\ \mathrm{d}x \!\ \frac{1}{p^2} x^{ip -1}
\left( f_{\pm}^{RR}(x,z)\right)^* \sin[\pi m (\frac{1}{2} + i p )]
\sin[\pi n (\frac{1}{2} + i p )].
\end{equation}
We are interested in computing the field strength out of these
integrals: thus we apply the operators $\mathcal{D}_{RR}^{\pm}$
(\ref{1orderfull}) to obtain an additional factor of $p$.
Moreover the operator (\ref{1orderfull}) changes the parity of
the function $f^{RR}_{\pm}(x,z)$, i.e. flips $f^{RR}_{\pm}(x,z)$, 
to  $f^{RR}_{\mp}(x,z)$.  We therefore need to consider, distinguishing
explicitly the two parities:
\begin{eqnarray}
\int \mathrm{d}p \!\ \mathrm{d}x &&\!\ \frac{p}{p^2} x^{ip -1}
f_{-}^{RR}(x,z) \times \\ 
&&\times \left\{ \begin{array}{ll} \sinh(\pi m p) \sinh(\pi n p)
 & \quad \mbox{if $m=$even, $n=$even} \\
\cosh(\pi m p) \cosh(\pi n p) & \quad \mbox{if $m=$odd, $n=$odd}
\end{array} \right.
\end{eqnarray}
and
\begin{eqnarray}
  \int \mathrm{d}p \!\ \mathrm{d}x &&\!\ \frac{p}{p^2} x^{ip -1}
f_{+}^{RR}(x,z) \times\\
&&\times \left\{ \begin{array}{ll} 
\sinh(\pi m p) \cosh(\pi n p) & \quad \mbox{if $m=$even, $n=$odd} \\
\cosh(\pi m p) \sinh(\pi n p) & \quad \mbox{if $m=$odd, $n=$even}
\end{array} \right.
\end{eqnarray}
Here the gamma-function squared in the numerator, 
as well as the  $p$-dependent phase, cancel 
with the normalization of the wave-function ${\Psi}^{RR}_{\pm}$:
thus the integral in $p$ is a (divergent) Gaussian integral that
need to be regulated and can be computed 
as we did in the mini-superspace approximation. 
Indeed, the $p$ and $\tau$ integrals are basically
the same as before: the factor
$e^{\pi pa}$ , in the notation of
section 3,  which before had $a=\pm 1$, now has generically
$a=\pm m \pm n$. Using the contour deformation procedure
already explained, one finds for the final $x$-integrals the
expressions, up to $m,n$ dependent phases and overall constants: 
\begin{eqnarray}\label{xinteven}
&& 
\left(2 \int_{e^{i\pi
 (m+n)}}^{1} 
-2 \int_{e^{i\pi
 (m-n)}}^{1}\right)\frac{1}{x}
f^{RR}_{-}(x,z).
\quad \mbox{if $m=$even,
 $n=$even} \cr
 &&\left( 2 \int_{1}^{\infty} 
+ 2 \int_{e^{i\pi
 (m+n)}}^{1} 
+ 2 \int_{e^{i\pi
 (m-n)}}^{1}\right) \frac{1}{x}
f^{RR}_{-}(x,z).
\quad \mbox{if $m=$odd,
 $n=$odd} 
\end{eqnarray}
The remaining two cases are:
\begin {eqnarray}\label{xintodd} 
&& 
\left(2 \int_{e^{i\pi
 (m+n)}}^{1} 
+2 \int_{e^{i\pi
 (m-n)}}^{1}\right) \frac{1}{x}f^{RR}_{+}(x,z)
\quad \mbox{if $m=$even,
 $n=$odd} \cr
 && \left(2 \int_{e^{i\pi
 (m+n)}}^{1} 
-2 \int_{e^{-i\pi
 (m-n)}}^{1}\right)\frac{1}{x}f^{RR}_{+}(x,z) 
\quad \mbox{if $m=$odd,
 $n=$even}
\end{eqnarray}
In both cases we  have used the properties:
\begin{eqnarray}
&& \int_0^1 \frac{\mathrm{d} x}{x} \!\ f^{RR}_{-}(x,z) = -
\int_1^{\infty} \frac{\mathrm{d} x}{x} \!\ f^{RR}_{-}(x,z) \cr &&
\int_0^1 \frac{\mathrm{d} x}{x} \!\ f^{RR}_{+}(x,z) =
\int_1^{\infty} \frac{\mathrm{d} x}{x} \!\ f^{RR}_{+}(x,z)
\end{eqnarray}
that follow from (\ref{xto1overx}).
Using (\ref{eqmot}) and the property obeyed by  $f^{RR}_{\pm}(x,z)$
given above, one can verify that the integrals
\eq{xinteven} and \eq{xintodd} are annihilated, in the $z\to 0$ limit,
by the operators ${\cal D}^-_{RR}$ and  ${\cal D}^+_{RR}$
respectively, i.e. that the equations of motion for the field
strength are satisfied. More generally, for arbitrary $z$, requiring
the equations of motion to be satisfied, will just put appropriate
conditions for the functions $f^{RR}_{\pm}(x,z)$ at the end points
of the integration contours in \eq{xinteven} and \eq{xintodd}.

One can also evaluate \eq{xinteven} and \eq{xintodd} 
in the limit $z\to 0$
using the asymptotic expression for $K_0(z)\sim -lnz$
\cite{Abram}, at least
for the integrals which involve a finite range of integration for
$x$. Let us start from the  case $(m,n)$ (even,even): 
Using the explicit form of $f^{RR}_-$ given above,
one finds that the relevant integral in (\ref{xinteven}) becomes
proportional to:
\begin{equation}
\left(\int_0^{\pi(m+n)}-\int_0^{\pi(m-n)}\right)
\mathrm{d} \theta\left(\ln z \sin{\frac{\theta}{2}}
+{\theta}\cos{\frac{\theta}{2}}\right),
\end{equation}
By doing the explicit $\theta$-integral one finds:
\begin{equation}
\left[-2\ln z \left((-)^{\frac{m+n}{2}}-1\right)+
4\left((-)^{\frac{m+n}{2}}-1\right)\right]-
\left[n\rightarrow -n\right],
\end{equation}
which vanishes for $(m,n)$ (even,even).
For the cases $(m,n)$ (even,odd),
using the asymptotic expression for $f_+^{RR}$  given in \eq{effes},
one obtains:
\begin{equation}
\left(\int_0^{\pi(m+n)}+\int_0^{\pi(m-n)}\right)
\mathrm{d} \theta\left(\ln z \cos{\frac{\theta}{2}}
-{\theta}\sin{\frac{\theta}{2}}\right).
\end{equation} 
Performing the $\theta$-integration as before, one finds:
\begin{equation}
\left[-2\ln z (-)^{\frac{m+n+1}{2}}+4(-)^{\frac{m+n+1}{2}}\right]+
\left[n\rightarrow -n\right],
\end{equation}
which vanishes.
Similarly for $(m,n)$ (odd,even) we have:
\begin{equation}
\left(\int_0^{\pi(m+n)}-\int_0^{\pi(m-n)}\right)
\mathrm{d} \theta\left(\ln z \cos{\frac{\theta}{2}}
+{\theta}\sin{\frac{\theta}{2}}\right),
\end{equation} 
which gives:
\begin{equation}
\left[-2\ln z (-)^{\frac{m+n+1}{2}}+4(-)^{\frac{m+n+1}{2}}\right]-
\left[n\rightarrow -n\right],
\end{equation}
which again vanishes.
Finally, we are left with the case $(m,n)$ (odd,odd). The corresponding
expression contains an integral which is independent of $m,n$, plus
two integrals that may depend on $(m,n)$ and can be evaluated, 
in the $z\to 0$ limit, as in the cases before. By an explicit 
calculation, one can indeed verify that, the $(m,n)$ dependence
drops out.   
The results we have found for the RR field, starting from the
``exact''wave-function in the $z\to 0$ limit, agree with
the mini-superspace analysis of section 3.

\subsection{ The NS-NS  field}

Let us now consider the NS case. The NS-NS field is proportional
to 
\begin{equation}
\int \mathrm{d}p \!\ \mathrm{d}x \!\ \frac{1}{p^2} x^{ip -1}
\left( f^{NS}(x,z)\right)^* p \sinh(\pi m p )
\sinh(\pi n p)
\Gamma^2(ip)(\frac{\mu}{2})^{-ip},
\end{equation}
where we have used $x\Gamma(x)=\Gamma(x+1)$.  The factor involving
the gamma-function above, as well as a $p$-
dependent phase  cancels, as usual, with the normalization
of $f^{NS}$. As a result, the integrand is an even function
of $p$, after taking into account the inversion  property
(\ref{xto1overx}) of $f^{NS}$. Proceeding as it was repeatedly
done before, we end up with an expression proportional to:
\begin{eqnarray}\label{nsfield1}
&&\left(\int_0^{\pi(m+n)}-\int_0^{\pi(m-n)}\right)
\mathrm{d}\theta\left( K_0\left(\sqrt{2zx}\right)-
K_0\left(\sqrt{\frac{2z}{x}}\right)\right)\nonumber,
\end{eqnarray}
where $x=e^{i\theta}$. Using again the behaviour $K_0(z)\sim -\ln z$ for
$z\to 0$, one finds that the expression (\ref{nsfield1}) becomes
\begin{equation}
\left(\int_0^{\pi(m+n)} -
\int_0^{\pi(m-n)}\right)\mathrm{d}\theta \theta
\end{equation}
which is proportional to $mn$. This is again in agreement
with the result of section 3. 
There the NS-NS tachyon field source by the 
$(m, n)$ brane is given by (\ref{nsfield}),
for  $z\rightarrow 0$ limit 
$I_0(z)\rightarrow 1$ 
therefore, this reduces to what we have found above.

\section{Probing ZZ with FZZT}

In this section we provide a further check on the  
results, obtained in the previous sections, 
about the field configurations produced by ZZ branes 
by using FZZT branes \cite{Fateev:2000ik,Teschner:2000md} 
as probes. The idea is thus to look at a
mixed, ZZ and FZZT annulus partition function, which
describes the exchange of closed string states between
ZZ and  FZZT branes, and extract from this
information about the NSNS or RR field produced by the ZZ
brane. 
The annulus partition function with mixed ZZ--FZZT boundary 
conditions for either the NSNS or RR closed string
sectors, $\mathcal{Z}^{NS,R}(m,n|\sigma)$, 
depends on the ZZ labels $(m,n)$, whose meaning
has been explained in section 2, and on a continuous parameter,
$\sigma$, for FZZT (Neumann) boundary conditions. 
This latter indeed characterizes 
continuous, non-degenerate representations of Liouville theory
and is related to the boundary cosmological constant $\mu_B$
through $\mu_B=\cosh\pi b\sigma$, for arbitrary $b$. 

\subsection{The NS-NS field}

Given  $\mathcal{Z}(m,n|\sigma)$ in the NSNS sector, 
following \cite{Kutasov:2004fg}, we introduce the
field $\Psi(z)$, encoding information on the target space
NSNS background, through:
\begin{equation} \label{partitionfunction}
\mathcal{Z}^{NS}(m,n | \sigma) = \int_{0}^{\infty} \frac {\mathrm{d} z}{z} \!\
e^{-\frac{z}{2}\left( x + \frac{1}{x} \right)} \Psi(z),
\end{equation}
where we have introduced the variable $x = e^{\pi \sigma}$
and set $b=1$. Notice that, in this case $x+1/x=\mu_B$, and
the definition  (\ref{partitionfunction}) is partly motivated 
by the the intuition from the mini-superspace approximation,
where one can see that $\mu_B$ is conjugate to $z$. Indeed,
in this approximation, the
wave function corresponding to a FZZT brane vanishes roughly for
$z$ larger than $\mu_B^{-1}$.
To evaluate (\ref{partitionfunction}) we will
compute $\mathcal{Z}^{NS}(\sigma' | \sigma)$ i.e. the annulus
amplitude corresponding to FZZT-FZZT boundary conditions,
and then analytically
continue in the parameter $\sigma$,  by using the 
known relation between FZZT and ZZ boundary states, which
amounts to the identity \cite{Martinec:2003ka}:
\begin{equation} \label{zzprescription}
\mathcal{Z}^{NS}\left(m,n | \sigma \right) = \mathcal{Z}^{NS}\left(
\sigma'_{m,-n}=i(m-n) | \sigma \right) - \mathcal{Z}^{NS}\left(
\sigma'_{m,n} = i (m+n) | \sigma \right).
\end{equation}
Similar relations hold in the RR sector.

Let us begin by evaluating the FZZT-FZZT partition function. 
One can think of it, for example,  as expressing the
interaction between a D-instanton and a (unstable and space-like ) 
D0-brane in 0B string theory, where for both it is assumed
a uniform distribution in the time direction,
as in section 3. Notice that the oscillator modes
cancel between matter and (super-)ghosts. Therefore,
it is given just in terms of the FZZT the boundary state wave functions 
$\Psi^{NS}(p,\sigma)$ as:
\begin{equation}\label{annulus}
\mathcal{Z}^{NS}(\sigma' | \sigma)=\int \mathrm{d}p(\Psi^{NS}(p,\sigma')^*
\frac{1}{p^2}\Psi^{NS}(p,\sigma),
\end{equation}
where, up to an overall constant:
\begin{equation}\label{neumannns}
\Psi^{NS}(p,\sigma)=i\cos(\pi \sigma p)\frac{\Gamma^2(1+ip)}{p}
(\frac{\mu}{2})^{-ip}.
\end{equation}
Using this, we arrive at:
\begin{equation}
\mathcal{Z}^{NS}(\sigma' | \sigma) 
= \int \mathrm{d} p \!\ \frac{\cos(p
\pi \sigma') \cos(p \pi \sigma)}{\sinh^2(\pi p)} \frac{1}{p^2}.
\end{equation}
To continue with the evaluation, we find it convenient to act with the
operator $x \partial_x=\frac{1}{\pi}\partial_\sigma$ 
on the identity (\ref{partitionfunction})
to lower the IR divergence in the momentum integration. 
The relevant integral is now:
\begin{equation}
\frac{1}{\pi} \frac{\partial}{\partial\sigma} 
\mathcal{Z}^{NS}(\sigma' | \sigma)  =
- \int \mathrm{d} p \!\ \frac{\cos(p \pi \sigma') \sin(p \pi
\sigma)}{\sinh^2(\pi p)} \frac{p}{p^2},
\end{equation}
which, in turn, equals the following integral:
\begin{equation}
\frac{1}{\pi} \frac{\partial}{\partial\sigma} 
\mathcal{Z}^{NS}(\sigma' | \sigma)  =
-\frac{1}{2 i} \int \mathrm{d} p \!\ \frac{e^{i p \pi (\sigma +
\sigma')}+ e^{i p \pi (\sigma - \sigma')}}{\sinh^2(\pi p)}
\frac{p}{p^2}.
\end{equation}
We will compute this integral by using the residue theorem; to do this,
we will think of $p$ as a complex variable and evaluate the poles
of the integrand. We will assume both $\sigma + \sigma'$ and
$\sigma - \sigma'$ positive and choose a contour that closes in
the upper half plane. At the end of the computation we will show
our results to be analytic in $\sigma$ and $\sigma'$. Particular
attention should be payed when evaluating the double pole at $p =
0$ since it lies on the integration contour: a defining
prescription is needed. The contribution from the poles gives (up to an
overall coefficient):
\begin{eqnarray} \label{NSpoles}
\frac{1}{\pi} \frac{\partial}{\partial\sigma} 
\mathcal{Z}^{NS}(\sigma' | \sigma)  
 &=&\frac{1}{\pi^2} \sum_{k=1}^{\infty} \frac{i \pi (\sigma +
\sigma') e^{-k \pi (\sigma+\sigma')} + i \pi (\sigma - \sigma')
e^{-k \pi (\sigma - \sigma')} } {i k}  \\ \nonumber
&+ & \frac{1}{\pi^2} \sum_{k=1}^{\infty} \frac{e^{-k \pi
(\sigma+\sigma')} + e^{-k \pi (\sigma - \sigma')} } {k^2}  \\
\nonumber 
&-& \frac{1}{4} (\sigma + \sigma')^2 - \frac{1}{4}
(\sigma - \sigma')^2 - \frac{1}{3}.
\end{eqnarray}
The structure of the above equation is as follows: the first two
lines are arising from the second order poles at $p = i k$, $k$
being a positive integer. The last line is due
to the pole at $p = 0$ for which a particular prescription is
needed. We chose to use the following $\epsilon$--prescription:
\begin{equation}
\frac{1}{\pi} \frac{\partial}{\partial\sigma} 
\mathcal{Z}^{NS}(\sigma' | \sigma)  = 
- \lim_{\epsilon\rightarrow 0}
\frac{\pi}{2 i} \int \mathrm{d} p \!\ \frac{e^{i p \pi (\sigma +
\sigma')}+ e^{i p \pi (\sigma - \sigma')}}{\sinh(\pi (p + i
\epsilon)) \sinh(\pi (p - i \epsilon))} \frac{p}{(p+i \epsilon)(p
- i \epsilon)}.
\end{equation}
Let us focus only on the first exponential in the above equation. 
The evaluation of the
$p = i \epsilon$ pole gives:
\begin{equation}
O\left(\frac{1}{\epsilon^2} \right) + O \left(\frac{1}{\epsilon}
\right) 
-\frac{1}{4} \left( \sigma + \sigma' \right)^2 - \frac{1}{6}
+ O(\epsilon) + \cdots
\end{equation}
Note that, as in \cite{Kutasov:2004fg} the 
the pole at $p -i\epsilon$ contains divergent pieces from 
which the finite part as $\epsilon \rightarrow 0$ has to be 
extracted.
This leads us to the final result of (\ref{NSpoles}). Now we
proceed in summing the series, by relying on the formulas
\begin{equation}
 \sum_{k=1}^{\infty} \frac{x^k}{k} = - \log(1-x),  \;\;\;\;\;\;\;\;
 \sum_{k=1}^{\infty} \frac{x^k}{k^2} = \mathrm{Li}_2(x),
\end{equation}
where $x$ is understood to lie in the convergence domain of the respective
sum. These sums allow us to rewrite (\ref{NSpoles}) as:
\begin{eqnarray} \label{sumNS}
\nonumber
\frac{1}{\pi} \frac{\partial}{\partial\sigma} 
\mathcal{Z}^{NS}(\sigma' | \sigma)  &=& 
- \frac{1}{\pi} (\sigma + \sigma') \log \left( 1 - e^{- \pi
(\sigma + \sigma')} \right) - \frac{1}{\pi} (\sigma - \sigma')
\log \left( 1 - e^{- \pi (\sigma - \sigma')} \right)  \\ \nonumber
&+& \frac{1}{\pi^2} \mathrm{Li}_2 \left(e^{- \pi (\sigma +
\sigma')} \right) + \frac{1}{\pi^2} \mathrm{Li}_2 \left(e^{- \pi
(\sigma - \sigma')} \right) \\  
&-& \frac{1}{4} (\sigma
+ \sigma')^2 - \frac{1}{4} (\sigma - \sigma')^2 -\frac{1}{3}.
\end{eqnarray}
This result holds for $\sigma + \sigma' > 0$ and $\sigma -
\sigma' > 0$. 

We now extend it to all values of $\sigma$ an $\sigma'$
by analytic continuation. For this purpose let us prove that eq.
(\ref{NSpoles}) is analytic in $\sigma+\sigma'$ and 
$\sigma-\sigma'$. Let us start with $\sigma+\sigma'$. 
Writing $\sigma$ for  $\sigma+\sigma'$, our result is of the form:
\begin{equation}
\label{gth0}
-\frac{\sigma}{\pi} \log(1 - e^{-\pi \sigma}) + \frac{1}{\pi^2}
\mathrm{Li}_2 (e^{-\pi \sigma}) - \frac{1}{4} \sigma^2
-\frac{1}{6},
\end{equation}
for $\sigma > 0$. On the other hand, for $\sigma < 0$ we have:
\begin{equation}
\label{lth0}
-\frac{\sigma}{\pi} \log(1 - e^{\pi \sigma}) - \frac{1}{\pi^2}
\mathrm{Li}_2 (e^{\pi \sigma}) + \frac{1}{4} \sigma^2 +\frac{1}{6},
\end{equation}
where an overall minus sign is due to the change of orientation in
the contour of integration, when picking up poles in the lower half
plane. Analyticity means that the two expressions 
are actually the same function defined in the complex 
 $\sigma$ plane. In other words, if we continue
the first expression to $\sigma<0$ we get the second
expression and vice-versa if we continue the second
expression to $\sigma >0$. 
This is easily proven by using the
following property of the di-logarithm function \cite{LEW}:
\begin{equation} \label{dilog}
\mathrm{Li}_2(y) + \mathrm{Li}_2(\frac{1}{y}) = \frac{\pi^2}{3} -
\frac{1}{2} \log^2 y - i \pi \log y,
\end{equation} 
that holds for $y > 1$. In our case it reads:
\begin{equation}
\mathrm{Li}_2(e^{-\pi \sigma}) + \mathrm{Li}_2(e^{\pi \sigma}) =
\frac{\pi^2}{3} - \frac{1}{2} \pi^2 \sigma^2 - i \pi^2 \sigma
\end{equation}
By using this, and the identity $\log(1-z)= i \pi +
\log(z-1)$, one verifies that 
the difference of \eq{gth0} and \eq{lth0} vanishes.
This ensures (\ref{sumNS}) is 
indeed analytic in $\sigma$. The same argument applies
to the dependence of (\ref{sumNS}) on the
 combination  $\sigma-\sigma'$.

Finally, we have to use the prescription (\ref{zzprescription}) on
the result (\ref{sumNS}). Let us suppose $m \pm n$ an even number.
Then, we can systematically replace factors like $e^{\pm i \pi (m
\pm n)}$ with $+1$. (for $m \pm n$ odd, we would have to replace
$e^{\pm i \pi (m \pm n)}$ with $-1$, but the results are unchanged). 
Then as a consequence of applying the identity
(\ref{zzprescription}) to (\ref{sumNS}), on obtains the final
result:
\begin{equation}
mn \vartheta(\sigma),
\end{equation}
where:
\begin{equation}
\label{step}
\vartheta (\sigma) = \left\{ \begin{array}{ll} 
+1 & \quad \mbox{if $\sigma > 0$} \\
-1 & \quad \mbox{if $\sigma<0$}
\end{array} \right.
\end{equation}
At this point one might wonder why we have obtained a step function
when \eq{sumNS} was shown to be analytic. This is because \eq{sumNS}
has branch cuts and when  the prescription
\eq{zzprescription} is applied one evaluates jumps across the branch
cuts \footnote{As a simple example consider $
\log(1-z) - \log(1-e^{2\pi i }z )$, this vanishes for $|z|\leq 1$ and
$2\pi i $ for $|z| >1$.}.
Finally, we can plug these results back in
(\ref{partitionfunction}) and, by recalling they were obtained after
applying the operator $x \partial_x$, we obtain:
\begin{equation}
mn \vartheta(\sigma) = {\cal K}  \int \frac {\mathrm{d} z}{z} \!\
(-\frac{z}{2}) \left( x - \frac{1}{x} \right)
e^{-\frac{z}{2}\left( x + \frac{1}{x} \right)} \Psi(z),
\end{equation}
or, equivalently:
\begin{equation}
\frac{mn \vartheta(\sigma)}{\sinh \sigma} = {\cal K}  \int \mathrm{d} z \!\
e^{-\frac{z}{2}\left( x + \frac{1}{x} \right)} \Psi(z).
\end{equation}
Here ${\cal K}$ stands for the overall constant we have not kept track
of.
By taking the inverse Laplace transform on the LHS 
using the tables in \cite{Grad} one finally ends
up with:
\begin{equation}
\Psi(z) =  m n I_0(z) \, {\cal K},
\end{equation}
which happily, agrees with the results of sections 3. and 4.

\subsection{The RR field}

In this section we extend the analysis of the previous subsection to
the RR case.  We introduce the target space field $\Psi_{\pm}^R$ 
which carries the information of the RR closed string field sourced by 
the D-instanton. These are related the annulus amplitude 
$\mathcal{Z}^R\left(m,n | \sigma \right)$  in the Ramond sector by the
following transforms
\be{defrrtrans1}
\mathcal{Z}^R\left(m,n | \sigma \right) = 
\int_0^\infty
\frac{dz}{\sqrt{z}} e^{-z\cosh\pi\sigma} \cosh \frac{\pi\sigma}{2} 
\Psi_+^{\rm RR} (z),
\ee
where $e^{\pi \sigma} =x$.
This definition of the target space field is the extension of the
definition of the target space field given in 
\eq{partitionfunction} for the Ramond
sector. It is motivated from the fact that the mini-superspace wave
$\Psi_{p+}(z)$ in \eq{mssfunctionsr} can be written 
in terms of a cosine using a Backlund transform
\cite{Douglas:2003up}, which is given by
\be{backlund}
\frac{\pi \cos p\pi\sigma}{\cosh \pi p} = \int_0^\infty
\frac{dz}{\sqrt{z}} e^{-z\cosh \pi \sigma} \Psi_{p+}(z) 
\ee
Applying $x\partial_x$ to the  equation \eq{defrrtrans1} we obtain
\bea{rrtrans}
x\frac{\partial}{\partial x} 
\mathcal{Z}^R\left(m,n | \sigma \right) &=& - 
\int_0^\infty \frac{dz}{\sqrt{z}} e^{-z\cosh\pi\sigma}
\sinh\frac{\pi\sigma}{2} 
\left( 
z \frac{\partial}{\partial z} + z \right) \Psi_+^{\rm RR} (z)
\\ \nonumber
&=& \int_0^\infty \frac{dz}{\sqrt{z}} e^{-z\cosh\pi\sigma}
\sinh\frac{\pi\sigma}{2}  \Psi_-^{\rm RR} (z)
\eea
In the first line of the above equation we have converted the
derivative $\partial_\sigma$ to a differential operator in $z$
using integration by parts. The
second line of the above equation is just the definition of
$\Psi_-^{\rm RR}(z)$.
To proceed 
we need the corresponding annulus amplitude in
the Ramond sector. 
This is obtained using the 
FZZT-FZZT annulus amplitude in the Ramond sector, which in 
turn is constructed from the 
RR boundary state wave-functions, $\Psi^R_{\pm}(p.\sigma)$, 
in (\ref{annulus}).
$\Psi^R_{\pm}(p.\sigma)$, come
in pair, and their expressions are proportional to:
\begin{equation}
\label{evenfn}
\Psi_+^{R}(p,\sigma)=\cos(\pi \sigma p){\Gamma^2(\frac{1}{2}+ip)}
(\frac{\mu}{2})^{-ip}.
\end{equation}
and:
\begin{equation}
\label{oddfn}
\Psi_-^{R}(p,\sigma)=\sin(\pi \sigma p){\Gamma^2(\frac{1}{2}+ip)}
(\frac{\mu}{2})^{-ip}.
\end{equation}

\vspace{.5cm}
\noindent
{\emph {Case i. $m \pm n$ even}}

Let us begin with the case of $m\pm n$ even for the ZZ boundary
state: we can obtain it from the FZZT boundary state $\Psi^R_+$,
and as a result the ZZ-FZZT amplitude can be written in terms
the FZZT-FZZT amplitude as: 
\begin{equation}
\label{rrzzpres}
\mathcal{Z}^R\left(m,n | \sigma \right) 
= \mathcal{Z}^R\left(
\sigma'_{m,-n}=i(m-n) | \sigma \right) \mp \mathcal{Z}^R\left(
\sigma'_{m,n} = i (m+n) | \sigma \right)
\end{equation}
where the $-$ sign is for both $m$ and $n$ even while the $+$ 
sign applies to the case of
both $m$ ad $n$ odd ($m \pm n$ is always even).
The FZZT-FZZT partition function is on the other hand:
\begin{equation}\label{annulusR}
\mathcal{Z}^R(\sigma' | \sigma) = 
\int \mathrm{d} p \!\ \frac{\cos(p
\pi \sigma') \cos(p \pi \sigma)}{\cosh^2(\pi p)} \frac{1}{p^2}.
\end{equation}
After applying the operator $x \partial_x$, $x=e^{\pi \sigma}$, one
finds:
\begin{equation}
x \frac{\partial}{\partial x} \mathcal{Z}^R(\sigma' | \sigma) = 
- \pi \int \mathrm{d} p \!\ \frac{\cos(p \pi \sigma') \sin(p \pi
\sigma)}{\cosh^2(\pi p)} \frac{p}{p^2},
\end{equation}
and the relevant integral to compute is:
\begin{equation}
\label{relint}
x \frac{\partial}{\partial x} \mathcal{Z}^R(\sigma' | \sigma) = 
-\frac{\pi}{2 i} \int \mathrm{d} p \!\ \frac{e^{i p \pi (\sigma +
\sigma')}+ e^{i p \pi (\sigma - \sigma')}}{\cosh^2(\pi p)}
\frac{1}{p}.
\end{equation}
Again, we evaluate this integral in the complex plane; to
do this we choose $\sigma + \sigma'$ and $\sigma - \sigma'$ to be
positive and close the contour in the upper half plane. At the
end we will show that
the expression we obtain will be
analytic in $\sigma$ and $\sigma'$.  The above
integral has double poles at $p = i (k + \frac{1}{2})$ and a
single pole at $p=0$ for which an $\epsilon$-prescription is
required. 
We use the following $\epsilon$ prescription 
on the integral \eq{relint}
\begin{equation}
x \frac{\partial}{\partial x} 
\mathcal{Z}^R(\sigma' | \sigma) = 
-\lim_{\epsilon \rightarrow 0} 
\frac{\pi}{2 i} \int \mathrm{d} p \!\ \frac{e^{i p \pi (\sigma +
\sigma')}+ e^{i p \pi (\sigma - \sigma')}}{\cosh^2(\pi p)}
\frac{p}{(p+i \epsilon)(p - i \epsilon)},
\end{equation}
By evaluating the residues at the poles, one finds:
\begin{eqnarray} \label{RRpoles}
 \nonumber
x \frac{\partial}{\partial x} \mathcal{Z}^R(\sigma' | \sigma) &=& 
 -\frac{1}{\pi^2} \sum_{k=0}^{\infty} \frac{i \pi (\sigma' +
\sigma) e^{- \pi ( k+\frac{1}{2} ) (\sigma + \sigma')} +i \pi
(\sigma - \sigma') e^{- \pi ( k+\frac{1}{2} ) (\sigma - \sigma')}
} {i (k + \frac{1}{2})}  
\\ 
&-&  \frac{1}{\pi^2}
\sum_{k=0}^{\infty} \frac{e^{- \pi ( k+\frac{1}{2} ) (\sigma +
\sigma')} + e^{- \pi ( k+\frac{1}{2} ) (\sigma - \sigma')} } { 
(k + \frac{1}{2})^2} + 1.
\end{eqnarray}
The last terms arises 
due to the pole at $p=i\epsilon$ 
which is given by 
\begin{equation}
\lim_{\epsilon \rightarrow 0} 
\frac{e^{ - \epsilon \pi (\sigma + \sigma')}+ e^{- \epsilon \pi
(\sigma - \sigma')}}{(\cosh \pi i \epsilon)^2} \frac{i \epsilon}{2
i \epsilon}  = \frac{1}{2} + \frac{1}{2} =  1.
\end{equation}
Now we resum the expressions in 
\eq{RRpoles} using the  following formulae:
\begin{equation}
 \sum_{k=0}^{\infty} \frac{x^{(k+\frac{1}{2})}}{k+\frac{1}{2}} =
- \log(1-\sqrt{x}) + \log (1 + \sqrt{x}), \;\;\;\;\;\;
 \sum_{k=0}^{\infty}
\frac{x^{(k+\frac{1}{2})}}{(k+\frac{1}{2})^2} = 2 \mathrm{Li}_2
(\sqrt{x}) - 2 \mathrm{Li}_2 (- \sqrt{x}).
\end{equation}
Using these formulae we rewrite equation (\ref{RRpoles}) as:
\begin{eqnarray}
\label{finrrpo}
x \frac{\partial}{\partial x} \mathcal{Z}^R(\sigma' | \sigma) &=& 
 - \frac{1}{\pi} 
 \left[  (\sigma + \sigma') \left( -
\log(1 - e^{-\frac{\pi}{2} (\sigma + \sigma')}) + \log(1 +
e^{-\frac{\pi}{2} (\sigma + \sigma')})
\right)   \right. 
\\ \nonumber &\; & \;\;\;\;\;\;\;\;\;\left. 
 + (\sigma - \sigma') \left( - \log(1 - e^{-\frac{\pi}{2} (\sigma
- \sigma')}) + \log(1 + e^{-\frac{\pi}{2} (\sigma -
\sigma')})\right) \right] 
\\ \nonumber 
&-&  \frac{2}{\pi^2} 
\left[ \mathrm{Li}_2 (e^{- \frac{\pi}{2} (\sigma + \sigma') }) -
 \mathrm{Li}_2 (- e^{- \frac{\pi}{2} (\sigma + \sigma') })  \right.
 \\ \nonumber
 &+& \;\;\;\;\;\;\;\;\left. 
  \mathrm{Li}_2 (e^{- \frac{\pi}{2} (\sigma - \sigma') }) -
    \mathrm{Li}_2 (- e^{- \frac{\pi}{2} (\sigma - \sigma') })
    \right] +1.
\end{eqnarray}

We now show that the expression in \eq{finrrpo} is an analytic
function of $\sigma + \sigma'$ and $\sigma - \sigma'$. 
Consider the terms in \eq{finrrpo} which dependent only on 
$\sigma + \sigma'$ and label this combination as 
$\sigma$, then these terms 
for $\sigma>0$ are given by
\begin{eqnarray}
\label{rrup}
&-& \frac{\sigma}{\pi^2}    \left( - \log(1 -
e^{-\frac{\pi}{2} \sigma }) + \log(1 + e^{-\frac{\pi}{2} \sigma
})\right) 
\\ \nonumber 
&-& 
\frac{2}{\pi^2} \left( \mathrm{Li}_2 (e^{- \frac{\pi}{2} \sigma }) -
 \mathrm{Li}_2 (- e^{- \frac{\pi}{2} \sigma}) \right) + \frac{1}{2}
\end{eqnarray}
whereas for $\sigma<0$ we have to close the contour in \eq{relint}
on the lower half plane, this gives:
\begin{eqnarray}
\label{rrdo}
&- &  \frac{\sigma }{\pi}  \left( - \log(1 -
e^{\frac{\pi}{2} \sigma }) + \log(1 + e^{\frac{\pi}{2} \sigma
})\right) 
\\ \nonumber 
&+& \frac{2}{\pi^2} \left( \mathrm{Li}_2 (e^{ \frac{\pi}{2} \sigma }) -
 \mathrm{Li}_2 (- e^{ \frac{\pi}{2} \sigma}) \right) - \frac{1}{2}
\end{eqnarray}
Here  an overall minus sign appears because of the change of
orientation in the integration contour and 
$\sigma$ in \eq{rrup} is replaced by $-\sigma$. 
To show is analyticity in $\sigma$ we have to show that the
difference of \eq{rrup} and \eq{rrdo} vanishes. On taking 
the difference we have to use the following identities satisfied by 
the di-logarithms.
\begin{equation}
\mathrm{Li}_2(e^{-\frac{\pi}{2} \sigma}) +
\mathrm{Li}_2(e^{\frac{\pi}{2} \sigma}) = \frac{\pi^2}{3} -
\frac{1}{2} \left(\frac{\pi\sigma}{2}\right)^2  + 
i \frac{\pi^2}{2} \sigma
\end{equation}
This identity is obtained from \eq{dilog} by setting 
$y = e^{\pi\sigma/2 }$. 
It is also convenient to use the following identity \cite{LEW}
\begin{equation}
\mathrm{Li}_2(-e^{-\frac{\pi}{2} \sigma}) +
\mathrm{Li}_2(-e^{\frac{\pi}{2} \sigma}) = 
- \frac{\pi^2}{6} - \frac{1}{2} \left( \frac{\pi \sigma}{2} \right)^2
\end{equation}
Using these identities one can indeed show that the 
the difference between \eq{rrup} and \eq{rrdo} vanishes. 
The same argument applies to the terms depending on the combination 
$\sigma -\sigma'$ in \eq{finrrpo}. Therefore we have shown that
the final expression in \eq{finrrpo} in an analytic function in 
the $\sigma+ \sigma'$ and $\sigma -\sigma'$. 

Now we have to use the prescription  given in \eq{rrzzpres}
to finally obtain the ZZ-FZZT annulus amplitude. Using the expression 
in \eq{finrrpo} for the 
FZZT-FZZT annulus amplitude in the Ramond sector we obtain
\bea{finrro}
\frac{\partial }{\partial\sigma}
\mathcal{Z}^R(\sigma' | \sigma) &=&  0 \;\;\;\;\;\;
m, n \;\;\;{\rm even } \\ \nonumber
&=& {\cal K} \, \vartheta (\sigma )\;\;\;\;\;\; m, n \;\;{\rm odd}
\eea
Here ${\cal K}$ refers to the overall normalization which we have
not kept track of and $\vartheta$ is the step function introduced in
\eq{step}.  
Using the definition of $\Psi_-^{\rm RR}$ given in \eq{rrtrans} and
taking the appropriate inverse Laplace transform we obtain
\bea{finrrfield}
\Psi_-^{\rm RR} &=& {\cal K} \, e^z \;\;\;\;\; m, n \;\;{\rm odd} , \\
\nonumber
&=& 0 \;\;\;\;\; m, n \;\;{\rm even}
\eea

\vspace{.5cm}
\noindent
{\emph {Case ii. $m\pm n$ odd}}

To obtain the ZZ-FZZT amplitude when 
when $m\pm n$ is odd we have to use the FZZT boundary state wave
function $\Psi_-^R(p, \sigma)$ given in \eq{oddfn}. 
The FZZT-FZZT partition function is then written as
\be{oddfzzt}
\mathcal{Z}^R(\sigma' | \sigma) = 
\int dp \frac{\sin (p\pi \sigma') \sin (p\pi \sigma) }{\cosh^2 (\pi p)
} \frac{1}{p^2}
\ee
Given the above partition function the ZZ-FZZT amplitude is given by 
the identity
\be{defzzfi}
\mathcal{Z}^R(m, n|\sigma) =
\mathcal{Z}^R( \sigma' = i(m+n) |\sigma) \mp 
\mathcal{Z}^R( \sigma' = i(m-n) |\sigma)  
\ee
where the $-$ sign is for $m$ odd and $n$ even while the $+$ sign
is for the case of $m$ even and $n$ odd. 
Now we proceed as before by first applying the operator 
$x\partial_x$ with $x=e^{\pi \sigma}$ on \eq{oddfzzt}, this gives
\be{diffzzzt}
\pi x \frac{\partial}{\partial x}
 \mathcal{Z}^R(\sigma' | \sigma) = 
\pi \int dp 
\frac{ \sin(p \pi \sigma') \cos ( p \pi \sigma) }{\cosh^2(\pi p)} 
\frac{p}{p^2}
\ee
We can now re-write the above integral in the convenient form given
below
\be{confzzt}
\pi x \frac{\partial}{\partial x}
 \mathcal{Z}^R(\sigma' | \sigma) = 
\frac{\pi}{2i} \int dp \frac{e^{ip \pi (\sigma + \sigma') } -
e^{ip \pi (\sigma-\sigma')} }{\cosh^2(\pi p) } \frac{1}{p}
\ee
Comparing the above expression with \eq{relint} we see that the only
change is the presence of the relative sign between the two terms. 
Therefore using the same steps as discussed for the case of
$m\pm n$ even, we obtain the following final formula for the 
ZZ-FZZT amplitude
\begin{eqnarray}
\label{finrrodd}
x \frac{\partial}{\partial x} 
\mathcal{Z}^R(\sigma' | \sigma) &=& 
 - \frac{1}{\pi} 
 \left[  (\sigma + \sigma') \left( -
\log(1 - e^{-\frac{\pi}{2} (\sigma + \sigma')}) + \log(1 +
e^{-\frac{\pi}{2} (\sigma + \sigma')})
\right)   \right. 
\\ \nonumber &\; & \;\;\;\;\;\;\;\;\;\left. 
 - (\sigma - \sigma') \left( - \log(1 - e^{-\frac{\pi}{2} (\sigma
- \sigma')}) + \log(1 + e^{-\frac{\pi}{2} (\sigma -
\sigma')})\right) \right] 
\\ \nonumber 
&-&  \frac{2}{\pi^2} 
\left[ \mathrm{Li}_2 (e^{- \frac{\pi}{2} (\sigma + \sigma') }) -
 \mathrm{Li}_2 (- e^{- \frac{\pi}{2} (\sigma + \sigma') }) \right.
 \\ \nonumber
 &-& \left. \;\;\;\;\;\;\;
  \mathrm{Li}_2 (e^{- \frac{\pi}{2} (\sigma - \sigma') }) +
    \mathrm{Li}_2 (- e^{- \frac{\pi}{2} (\sigma - \sigma') })
    \right] .
\end{eqnarray}
From the arguments discussed for the previous case it is clear that
the above expression is an analytic function of $\sigma + \sigma'$ and
$\sigma -\sigma'$. 
To obtain the ZZ-FZZT amplitude we can now use the prescription given
in \eq{defzzfi}. Substituting the expression \eq{finrrodd}  
for the ZZ-FZZT amplitude we see that the 
\be{fullform}
x\frac{\partial}{\partial x } 
\mathcal{Z}^R(m, n | \sigma) =  0 , \;\;\;\;\; (m \pm n ) \;\; 
{\rm odd}
\ee

Therefore we obtain the result that the RR charge by a $(m,n)$ D-instanton
vanishes unless $m,n$ are both odd and is equal to that of the
$(1,1)$ D-instanton.

\section{Conclusions}

In this paper we have used three methods to study the behaviour of
the closed string fields $\Psi (z) $ sourced by $(m,n)$ D-instantons
of type 0B theory distributed uniformly in the time direction. 
The three approaches were:  the mini-superspace
method, the extension of wave-functions beyond the mini-superspace
approximation and finally using the ZZ-brane as a probe of the 
closed string field. 
In all the three methods we obtained 
the tension of an $(m,n)=(t,1)$ brane is proportional
to 
$mn=t$ times the tension of the $(1,1)$ brane and the RR scalar charge 
of these branes is non-zero only for the case of $t$ odd 
and is equal to that of the $(1,1)$ brane. 
As a further consistency check, we verified that the closed string
fields sourced by the D-instantons satisfied the corresponding 
equations of motions.
The consistency of these results 
lends support to the three methods.
In fact we find the functional dependence of the closed
field from the mini-superspace approach and that determined 
by using the ZZ-probe method agrees identically. 
It will be interesting to provide a  reason for this, because in
principle the exact closed string 
wave-functions and the  equations of motion can  be different
from the mini-superspace limit.
These results also imply that 
that the $(t,1)$ D0-branes of type 0A theory have the same property
as these are T-dual to the corresponding D-instantons considered here. 
These branes will carry RR one form charge of type 0A if
$t$ is odd and will be equal to that 
of the $(1,1)$ D0-brane.
At this point it is worth to make a comment about the relation 
between  our results and a result obtained in \cite{Seiberg:2003nm}.
There, the minimal $(p,q)$ string theory was studied, for which 
$b^2=p/q$ is rational. In this case the inequivalent  
ZZ boundary states are labelled by $(t-1,m,n)$ with $t\geq 1$,
$0<n\leq p$  and   $0<m\leq q$.  In   \cite{Seiberg:2003nm}
the boundary state corresponding to $(t-1,m,n)$ was shown to
be $t$ times the boundary state corresponding to $(0,m,n)$.
Setting formally $p=q=1$, so that $m=n=1$, to recover the 
$c=1$ case, this gives  the result that the tension
of $(t-1,1,1)$, which coincides with the state we
labelled by $(t,1)$,  goes like $t$, 
which formally agrees with our
results. However the result of   \cite{Seiberg:2003nm} is a consequence 
of a peculiar shift symmetry $\sigma\rightarrow \sigma +2i\sqrt{pq}$
in the FZZT boundary states which are used to construct the ZZ
boundary states. This is due to the fact that in the minimal
string case the closed string states in the BRST cohomology
involve only discrete Liouville states. This is to be contrasted with 
our case, where the matter is  a $c=1$ system and the BRST cohomology
involves also continuum, non-degenerate Liouville states, so that
the above symmetry is not obvious.  

The fact the tension of the $(t,1)$ brane increases 
as $t$ and the RR scalar charge
is just unity or zero indicates that these are unstable. 
This is in accordance with the presence of many tachyons in the open
string spectrum of these branes. From examining the annulus amplitude
of two $(t,1)$ D-instantons  with the appropriate GSO projection for
type 0B theory it is easy to see that there are  roughly 
of the order of $t$ tachyons \footnote{
For generic value of $b$ and $m>n$, the precise  number 
of tachyons is $mn -n$ if $2n\leq m$ and $mn -3n +m$ if $2n>m$;
for $m=n$ the number of tachyons is $mn-n$.}.
It is an open question to study the condensation of these tachyons, 
the height of the potential we now know is proportional 
to $t$. 
In the matrix formulation of type 0B theory the $(1,1)$ 
ZZ brane plays a special
role and is believed to be fundamental. 
Since the $(t,1)$ branes are not stable for $t>1$,  
it will be interesting to see if they 
can be thought 
of as  excited states
made of $(1,1)$ and anti-$(1,1)$ branes.  Finally, 
the $(m,n)$ ZZ boundary state can be constructed  for all values of $b$, 
in this paper we studied the $b\rightarrow 0$ limit and 
the point $b=1$. 
We have seen that 
for both these cases the tension and RR charge of these branes 
were same. 
It will be of interest to see if the results of 
the tension and the RR scalar charge of these branes
depend on the parameter $b$

\acknowledgments

We wish to thank Al. Zamolodchikov for a stimulating discussion and
useful comments, we are also grateful to Samir Murthy
for useful comments. 
M.C. also thanks B\'en\'edicte Ponsot for discussions.  
We thank N. Seiberg for bringing to our attention reference 
\cite{Seiberg:2003nm}, where degenerate  representations
and corresponding boundary states for $b^2$ rational are discussed.
The work of the authors is 
partially supported  by the RTN European programme:
MRTN-CT-2004-503369. 
The work of M.C. was supported in 
part by PPARC grant PPA/G/S/2002/00478

\bibliographystyle{utphys}
\bibliography{liouville}

\end{document}